\documentclass{aa}  

\usepackage{graphicx}
\usepackage{import}
\usepackage{newtxtext,newtxmath}
\usepackage[dvipsnames]{xcolor}
\usepackage{svg}
\usepackage{pgf}
\usepackage{comment}
\usepackage{soul}

\DeclareFontEncoding{OT2}{}{} 
\DeclareSymbolFont{cyrletters}{OT2}{cmr}{m}{n}
\DeclareMathSymbol{\Sha}{\mathalpha}{cyrletters}{"58} 

\newcommand{\chapr}{\Sha_{\rm Prot}}
\newcommand{\chafr}{\Sha_{\rm 1/Prot}}
\newcommand{\transit}{F_{\rm transit}}
\newcommand{\evol}{E_{\rm life}}
\newcommand{\life}{\tau_{\rm life}}
\newcommand{\ttransit}{\tau_{\rm tr}}
\newcommand{\prot}{ \rm P_{\rm rot}}
\newcommand{\teff}{\rm T_{\rm eff}}
\newcommand{\surf}{\rm A( \rm T_{\rm spot}, \rm S_{ \rm spot})}
\newcommand{\lnH}{\mathcal{H}}

\newcommand{\correct}[1]{{}}

\begin{document}

%

   \title{Activity of low mass stars in the light of spot signature in the Fourier domain}


  \author{L. Degott
          \inst{1,} \inst{2},
          F. Baudin
          \inst{1},
          R. Samadi
          \inst{2},
          B. Perri
          \inst{3}
          C. Pinçon
          \inst{1}
          }
\institute{Institut d'Astrophysique Spatiale (IAS),  CNRS, Université Paris-Saclay,
            Orsay,91400, France\\
        \email{lucie.degott@universite-paris-saclay.fr}
       \and
           LESIA, CNRS, Université Pierre et Marie Curie, Université Denis Diderot, Observatoire de Paris, 92195 Meudon Cedex, France
        \and
        Université Paris-Saclay, Université Paris Cité, CEA, CNRS, AIM, 91191, Gif-sur-Yvette, France
    }
   \date{Received 21 june 2024; accepted 17 february 2025}
   \titlerunning{Activity of low mass stars}
   \authorrunning{L. Degott et al.}
 
  \abstract
    {
    Magnetic fields exhibit a wide variety of behaviours in low mass stars and further characterization is required to understand these observations.
    Stellar photometry from space missions such as MOST, CoRoT, \textit{Kepler},\correct{TESS} and, in \correct{the near} future PLATO, provide thousands of highly precise light curves (LC) that can shed new light upon stellar activity, in particular through the signature of transiting spots.   
    }
   {We study the impact of star spots on light curves in the Fourier domain, reducing the degeneracies encountered by direct spot modeling in the temporal domain, and use this new formulation to explore the spot properties from the available data.
   
   }
   {We propose a model of LC power spectra at low frequency based on a description of spot transits that allows us to retrieve information about the amplitude of their photometric impact $\lnH$, and about the spot mean lifetime over the observation $\life$ when the power spectrum exibits rotation peaks .
   We first validate this method with simulated LCs and then apply it to the \textit{Kepler} data to extract global trends over a set of more than 37\, 755 stars.
   }
   {This analysis leads to a classification of the sample into “peakless” or “with peaks” spectra, and enables the identification of different activity regimes based on $\lnH$ and $\life$ for different ranges of Rossby number. More specifically, we observe an intense regime of activity between $\rm Ro = 0.7$ and $ \rm Ro = 1$, for stars with masses under $1 \rm M_{\odot}$.
   }
   {This new systematic method can be used to provide new observational constraints on stellar activity (and possibly a link with stellar magnetism)  when applied to large photometric datasets, such as those from the future PLATO mission.  
   }
 
    \maketitle


\section{Introduction}

Recent observations have revealed a diverse landscape for the magnetism of low mass stars.
This diversity is shown in the properties of their large-scale magnetic fields, which can influence not only the small-scale surface magnetic phenomena (spots, flares, faculae, etc…) but also the physical processes that govern the stars' interior and their environments. 
For example, magnetic fields are involved in \correct{the transport of energy and angular momentum} in stellar interiors by constraining the convection processes \citep{Proctor1982}. 
It has also been shown that the magnetic field can play a role in the star-exoplanet interactions by coupling their two fields \citep{Lanza2009, Lanza2013, Saur2013, Strugarek2015} and influence the activity of the host star \citep{Cuntz2000, Shkolnik2005}. 
The stellar wind and flares can also strongly affect the habitability of exoplanets through different processes of evaporation \citep{Vidal2004, Owen2019}. 
Understanding and characterizing this stellar magnetism is thus a crucial open question in astrophysics, since the magnetic field influences many phenomena related to stars and their environments.

To better understand the magnetism of stars, studying the Sun is especially convenient due to its proximity, which enables highly-detailed observations.  
Shortly after the Zeeman effect revealed the presence of magnetic field on the solar surface \citep{1908Hale}, \cite{1919Larmor} suggested that it could be explained by the dynamo effect, which is the capacity of a magnetized fluid to maintain and amplify its own magnetic field against Ohmic dissipation \citep{Moffatt1978, Parker1993}. 
There are several phenomena involved in the dynamo loop such as the differential rotation (DR), the turbulence in the convective envelope, the Coriolis force and/or the meridional circulation \citep{Brun2017, Charbonneau2020}. Depending on the stellar properties, these phenomena play various roles in the star's dynamo. 
A key parameter in the dynamo description is the Rossby number, which measures the effect of the convection compared to the rotation 
(see Sect. \,\ref{sc:results_kp} for a precise definition).
In particular, the dynamo effect helps to explain the Sun's 11-year sunspot cycle and the evolution of its magnetic topology, from dipolar configurations at minimum of activity to quadrupolar at maximum, leading to the inversion of its magnetic poles, with a periodicity of 22 years.

The presence of large-scale magnetic fields suggests that the dynamo effect is also at play inside other stars. However, the diversity and complexity of topologies observed is probably an indicator of different internal properties.
Some fields are much stronger than the solar one, while others are predominantly non-axisymmetric and toroidal compared to the Sun which has a more poloidal and axisymmetric field \citep{Donati2009}. 
The activity cycles also are extremely varied, with some stars showing cycles much shorter than the solar one, while others seem to have much longer cycles \citep{Baliunas1985, Jeffers2023}.   
\cite{Saar1999} observed that these stars are divided into two branches: one for stars with long cycles and slow rotation, called the “inactive branch”, and another for stars with faster rotation than the Sun and shorter cycles, called the “active branch”. 
Later, \cite{BV2007} linked these two regimes to different dynamo behaviours driven by different internal regions of the stars. 
Indeed, numerical simulations that reproduce similar activity cycles and other magnetic behaviours give us hints about the fact that the internal structures and differential rotation profiles of low mass stars are very diverse \citep{2022Brun}, reinforcing the idea that stellar dynamo behaviour is varied.
This makes it clear that the various ingredients of stellar dynamos need to be better characterized through more observational constraints.

As a consequence of stellar dynamos, number of magnetic phenomena appear on the stellar surface. This contributes to what we call stellar magnetic activity, which encompasses all the phenomena that induce luminosity variations in the signal produced by a star.
The phenomenon of interest in this paper is the activity due to the appearance of spots on stellar surfaces (hereafter “activity” for “spot activity”), which may be caused by the emergence of a magnetic structure at the surface, which locally freezes convection and prevents heat from rising up. This cools down locally these regions and makes them appear darker than the rest of the star's surface \citep{Solanki2003}.
These spots are a good tracer of the magnetic field because they are easy to observe and their properties (number, area, position, lifetime, etc…) are linked to the magnetic cycle.
Spots on our Sun have been observed and studied for centuries, however the knowledge gained from the behaviour of sunspots may not directly apply to other stars. The effectiveness of the solar analogy for more active stars is uncertain, especially considering that they frequently exhibit larger spotted areas with potentially longer lifetimes \citep{Valio2017, Namekata2019}.
Moreover, in the case of fast rotating stars the spots can be localized at the poles of the star, contrary to what we observe on the Sun \citep{2016Roettenbacher}.

A method to study these stellar spots is the use of precise photometric data. Since stars are rotating, dark spots appearing on their surface induce photometric modulations, which vary in time and amplitude depending on the physical properties of the spots. These variations are present in the light curves (LCs) from the
missions such as CoRoT \citep{Baglin2006}, \textit{Kepler} \citep{Borucki2010},TESS \citep{Ricker2014} and soon PLATO \citep{Rauer2014} that provide high precision photometric data for long periods \correct{of time }for thousands of stars, and thus allow us to retrieve information about spots for stars with different physical properties and work on a large data sample. 
In particular, the \textit{Kepler} mission has collected around 100 000 LCs from main-sequence stars over a period of 4 years.  

There are several indicators that can be used to study the activity of stars, such as the \ion{Ca}{ii} H and K lines \citep{Wilson1978, Noyes1984} emitted by the chromosphere or measurement of X-ray luminosity \citep{Vaina1981}. 
In photometry,  the $S_{ph}$ index \citep[see for ex.][]{Garcia2010, Mathur2014} is an indicator that allows us to follow magnetically induced modulations.
More recently, \cite{2022Basri} presented a method to retrieve mean spot lifetimes by using an autocorrelation function.
Another way to extract more precise information about spot properties by using LC is to fit these data in the temporal domain with a physical model by spot modelling \citep[see for ex.][]{Mosser2009}. 
Unfortunately, this method has proven to be complex to implement due to the numerous degeneracies of the problem. 
New methods of analysis could help to answer the remaining open questions about stellar magnetism.
The idea in this paper is to try to avoid some of these degeneracies encountered in the analysis of temporal time series by shifting to the Fourier domain, in which it is much easier to identify global trends about spot properties.
The better scale separation by sorting temporal frequency in the Fourier domain brings out average information about spots (lifetimes, transit time and coverage surface). For example, the periodic transit of spots can induce rotation period harmonics that are called “rotation peaks”. 

The impact of activity in the Fourier domain was first studied by \citet{Harvey85} and \citet{Harvey93} for the solar case by a heuristic method. 
Since the different modulations produced by spots granulation (convective cells visible in the photosphere), \correct{and super-granulation} occur on different timescales and can be seen as memory noises \correct{(i.e. the autocorrelation of the time series is a decreasing exponential)}, they can each be fitted by a pseudo-Lorentzian with a different width (see Sect\. ,\ref{sec2}). The spot activity having the longer timescale, it is located at the lower end of the spectrum.   

Later, \cite{Aigrain2004} showed that some characteristics of this pseudo-Lorentzian could be related to properties of the activity. 
More specific properties have also been studied: \citet{2017Santos} have studied the ratio between rotation peaks harmonics and demonstrated that it can be linked to the spot latitude and eventually to the inclination of the star. 

However, it is possible to go beyond these approaches and use all of the information contained in the power spectra to their full capacity. The aim of this work is to provide a more thorough method and apply it to a large dataset to build a landscape of stellar activity based on spot properties. 

We present here a new model that allows a more appropriate fit to the power spectrum of photometric data including the rotational peaks in order to extract global properties of the spots such as their lifetime and their photometric impact which is related to the spot area and the spot temperature contrast.
We begin by presenting this new model built on an analytical description of  a spot \correct{signature in a LC} in Section\,\ref{sec2}, following by a validation based on simulated LCs in Section \ref{sec:simu}.
We then detail our results by applying this method to the \textit{Kepler} data in Section\,\ref{sec:result} to take advantage of a large data set, which allows identifying trends and patterns. We then discuss this results in Section\,\ref{sec:disussion} and conclude these study in Section\,\ref{sec:conclusion}. 

\section{Modelling stellar spots in the Fourier domain} 
\label{sec2}

This section describes the analytical model proposed in this paper. 
This description is inspired by the original study of \cite{Harvey85}, \cite{Harvey93} and \citet{Aigrain2004}, which have previously worked in the Fourier domain to study the activity of the Sun. To describe the shape of the Fourier power spectra density (PSD), they used a pseudo-Lorentzian expression, which is: 

\begin{equation}
    PSD(\nu) = \frac{\mathcal{H}}{1 + (2 \rm \pi \tau \nu)^\alpha} \, ,
    \label{eq:Harvey}
\end{equation}
where $\nu$ is the frequency, $\lnH$ is the height of the Lorentzian, $\tau$ the timescale of the activity and $\alpha$ the exponent \citep[see][for the demonstration]{Harvey85}.
\cite{Aigrain2004} have linked certain properties of this pseudo-Lorentzian to solar activity features: they showed that $\lnH$ is related to the level of activity of the Sun, and suggested that $\tau$ could be related to the spot evolution timescale (of the order of a few days). This model has been widely used in stellar photometry to fit the contributions of magnetic activity or granulation. It has also been extended to complementary fields such as radial velocity observations or asteroseismology \citep[see for example][]{2009Michel,2012Appourchaux}.
However, this depiction overlooks certain properties of the spectrum, such as rotational peaks, suggesting that the model can be further improved, as proposed in this section.

We start by discussing the signature of a single spot case in photometry in the temporal domain and then in the Fourier domain in Sect.\,\ref{subsec:single_spot}.
Then we present the prescriptions used for the various functions describing the spot's evolution in Sect.\,\ref{sec:gaus_decription}. We present the general case with an ensemble of $N$ spots in Sect.\,\ref{subsec:n_spots}, and finally summarise the main parameters of our model in Sect.\,\ref{sec:model_conclusion}.
For the rest of the paper, we will use the following notation for the Fourier transform : 
\begin{equation*}
    TF[f(t)](\nu) = \int_{- \infty}^{+\infty} f(t) \rm e^{2 \rm i \rm \pi \nu t} \rm d t = \widehat{f}(\nu) = \widehat{f(t)}.
\end{equation*}

\subsection{Single-spot case}
\label{subsec:single_spot}

We consider a single stellar spot transiting a star with a rotation period  $\prot$, 
characterized by an area $\rm S_{ \rm spot}$ and a temperature $\rm T_{\rm spot}$. This spot is assumed to always rotate with the same rotation period and to remain at the same latitude. 
In this model, we will not consider the impact of differential rotation or faculae. The scope of this hypothesis will be discussed later in Section\,\ref{sec:facula_dr}. 

\begin{figure*}
   \resizebox{\hsize}{!}
            {\includegraphics[]{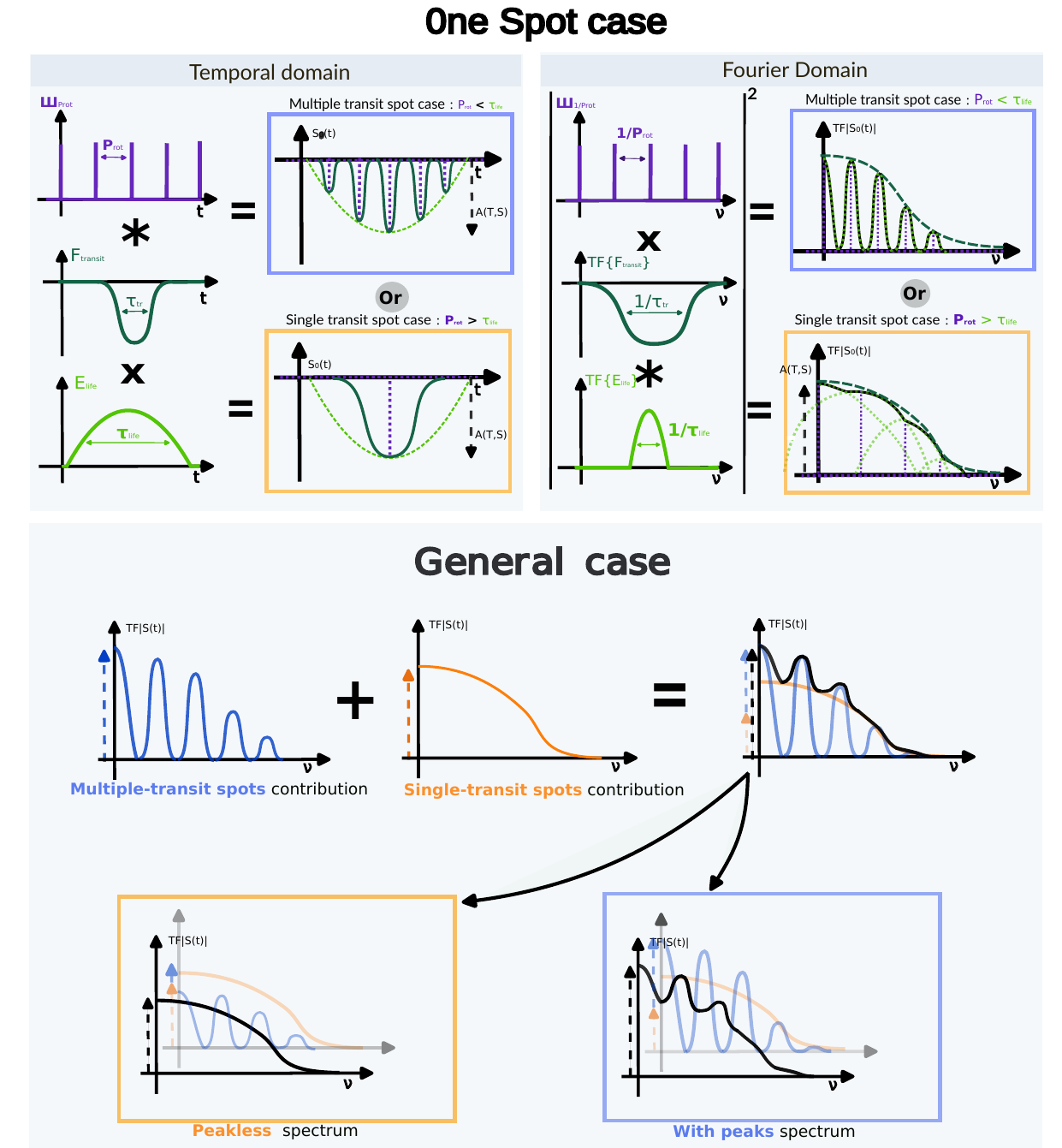}}
      \caption{
      Illustration of the analytical description of the spot transit proposed in this paper. 
      The top panel shows the model for the single-spot case. The model is shown in the time domain on the left, and in the Fourier domain on the right. Each time, the distinction is made between the single-transit-spot/peakless case (orange) and the multiple-transit-spot/with peaks case (blue). 
      The bottom of the figure illustrates the general case with both multiple-transit and single-transit spots' contributions.}
   
         \label{fig:model_ana}
   \end{figure*}

The transit of the spot on the stellar disc can be described using three functions, which are illustrated in~Fig. \ref{fig:model_ana}:

\begin{itemize}
      \item $\chapr$: a normalized Dirac comb with a spacing of one rotation period, which represents the periodicity of the transit.
      \item $\transit(t)$: a function to describe the mean shape of the transit.
      The shape of the transit can vary depending on the star's rotation axis relative to the line of sight, the position of the spot on the disc and limb-darkening effects.
      By considering that the transit time is defined by the width at mid-height,  noted $\ttransit$, its value is close to $\prot/4$, modulated by a factor depending on these configurations. 
      We consider here an average transit shape for the sake of simplicity.
      \item $\evol(t)$: a function to describe the intrinsic temporal evolution of the spot, with a characteristic time $\life$, which represents the spot lifetime. This function reflects the possible variations of the size and temperature of the spot during its lifetime.
   \end{itemize}
   
By combining these three components, the impact of the spot on the light curve with a unitary maximum depth, noted $S_0(t)$, can be reproduced. 
The convolution of $\chapr$ and $\transit(t)$ replicates the spot transits. Then, by multiplying by $\evol(t)$, we add the modulations induced by the intrinsic evolution of the spot. This yields for the light curve:

\begin{equation}
     S_0(t) = [\chapr * \transit(t)] \times \evol (t).
     \label{eq_time}
\end{equation}  

The value of the maximum transit photometric depth is given by $\surf$ which depends on the temperature contrast between the spot and the stellar surface, and the area of the spot.
Therefore, the total impact of a spot on a LC can be expressed as $S(t) = \surf \cdot S_0(t)  $. This model is illustrated in the top panel of Fig.\,\ref{fig:model_ana}. 

Two types of spectrum can be distinguished:
\begin{itemize}
    \item Stars where $\life$ > $\prot$: multiple transits of the spot are observed in the light curve; it is then categorized as the \textbf{multiple-transit-spot} case.
    \item  Stars where $\life$ < $\prot$: only one transit of the spot is visible in the light curve; it is then the \textbf{single-transit-spot} case.
\end{itemize}

When shifting to the Fourier domain, we obtain the following. 

\begin{equation}    
\widehat{S_0}(\nu)   =  \left[\frac{\chafr}{\prot} \times \widehat{\transit}\right] * \widehat{\evol}(\nu).
                \label{eq:S0}
\end{equation}
The convolution between the Fourier comb $\chafr$ and the Fourier transform of $\evol$ generates the rotation peaks that are seen in the light curve spectrum (modulations of the black line in the bottom panel of Fig. \ref{fig:model_ana}). These peaks are multiplied by the Fourier transform of the transit component, $\widehat{\transit}$, giving the final shape of the spectrum. 
We can then link the spot properties in the temporal domain to the spectrum properties in the Fourier domain, as summarised in Fig. \ref{fig:model_ana}. The information about the spot lifetime $\life$ appears thus in the rotation peaks. 
Depending on the spot lifetime, the transit component and the rotation harmonics will not appear the same.
In a multiple-transit-spot case, we have $\life$>$\prot$, and the height of the rotation peaks as a function of the frequency is given by the transit shape (see left side of Fig.\,\ref{fig:model_ana}). This can explain the results from \citet{2017Santos}, where they find that the peak ratio between the first and second \correct{rotational peaks} depends essentially on the transit profile of a single-spot in a LC. 
In the single-transit-spot case, we have $\life$<$\prot$, the rotation peaks widen  and overlap each other. As a result, the rotation peaks disappear and the spectrum comes close to the description of \citet{Harvey93}  for the solar case (see right side of Fig.\,\ref{fig:model_ana}). The information between $\life$ and $\ttransit$ is thus degenerated. 

\subsection{Gaussian case for the spot evolution} \label{sec:gaus_decription}

As a first step, we use Gaussians to describe $\evol$ and $\transit$. This description will be used later in Section\,\ref{sec:simu} for the simulated LC. 
These simple descriptions do not include the complexity of real star spots \correct{(spot shape, spot penombra, group of spot, etc...)}, but allows instead an analytical approach in order to understand how the modelled spectra evolve when spots characteristics are modified.
Indeed, similarly to \citet{Mosser2009}, a Gaussian function can be used to describe the evolution of the spot contrast over time: 
\begin{equation}
\evol(t) = \exp\left({-\frac{4 \ln 2 \cdot t^2}{\life^2}}\right) \, ,
\label{eq:mosser_description}
\end{equation}
where $\life$ is the characteristic lifetime of the spot seen as the width at half-height.

$\transit$ is described using the same Gaussian function :
\begin{equation}
\transit(t) = \exp\left({-\frac{4 \ln 2 \cdot t^2}{\ttransit^2}}\right).
\end{equation}

In the Fourier domain, and using Equation\,\ref{eq:S0}, we have: 
\begin{eqnarray}    
\widehat{S_0}(\nu)  & = & [\frac{\chafr}{\prot} \times{\frac{ \rm \pi}{\ln{2}2}}\frac{\ttransit}{2} \rm e^{-\frac{( \rm \pi \nu \ttransit)^2}{4 \ln{2}}} * \frac{\life}{2} \rm e^{-\frac{( \rm \pi \nu \life)^2}{4 \ln{2}}} \\
                & = & \frac{ \rm \pi}{4 \ln{2}} \life \ttransit [\frac{\chafr}{\prot}\times \rm e^{-\frac{( \rm \pi \nu \ttransit)^2}{4\ln{2}}}] * \rm e^{-\frac{( \rm \pi \nu \life)^2}{4\ln{2}}} \\
                & = & B \cdot [ \chafr\times \rm e^{-\frac{( \rm \pi \nu \ttransit)^2}{4 \ln{2}}}] * \rm  e^{-\frac{( \rm \pi \nu \life)^2}{4 \ln{2}}}\, ,
                \label{eq:gaus_approx}
\end{eqnarray}

where $B = \frac{ \rm \pi \ttransit}{4ln2 \prot} \life $ is a constant proportional to both the lifetime, $\life$, and the ratio $\ttransit / \prot$, which is correlated with the inclination of the stars. 

Knowing that \correct{the observed} inclination $i$ follows a probability distribution proportional to $\sin i$, we can assume that the majority of stars from the \textit{Kepler} mission that we will use in this paper have inclinations much closer to 90° than 0° (inclination axis parallel to the line of sight). The detection methods used to find the rotation periods are sensitive to the modulations generated by the spots, which are less and less visible when the star inclination is low (see Sect.\ref{subsec:kepler_data}).
This assertion should be treated with caution, as \cite{2023Bowler} found inclinations of down to 3.6° for stars whose rotation had been measured photometrically by periodograms, a method similar to the one used by \cite{Santos2019, Santos2021} (see Section\,\ref{subsec:kepler_data} for more details). However, the very small inclinations found in this paper remain a minority: 13$\%$ of these stars correspond to inclinations below 30°, and more than 50$\%$ above 60°. Assuming that the majority of stars \correct{of the sample use in this paper }have a high inclination, we have  $\ttransit \sim \frac{\prot}{4}$ and $B \sim \frac{ \rm \pi}{16 \ln{2} } \life $.

Other functions can be used to describe $\transit$ and $\evol$, that lead to the same physical interpretations. We will, for example, use Lorentzians or pseudo-Lorentzians in the Fourier domain when applying the model to the \textit{Kepler} data in Section\,\ref{sec:result}, since their shape appears to be more adapted to the observed spectra. 

\subsection{General case}
\label{subsec:n_spots}

When considering a light curve including $N$  spots (with $N \gg 1$), each one associated to a maximum depth $A_i(T_i, S_i)$ and a random appearance time $t_i$, the light curve  can be described as:

\begin{equation}
    S(t) = \sum_{i = 1}^{N} A_i \cdot S_0(t - t_i).
    \label{eq : white_noise}
 \end{equation}
In the Fourier domain, this yields: 

\begin{eqnarray}
      \widehat{S}(\nu) & = & \sum_{i = 1}^{N} A_i \cdot \widehat{S_0(t - t_i)}, \\
                    & = & \sum_{i = 1}^{N} A_i \cdot \rm e^{2\rm i \rm \pi \nu t_i}\widehat{S_0}(\nu)\,.
                  \label{eq:final1}
   \end{eqnarray}

This yields the following power spectra density (PSD) :
\begin{equation}
     |\widehat{S}(\nu)|^2  =  |A_{tot}|^2 \cdot |\widehat{S_0}(\nu)|^2 \, ,
    \end{equation}
with : 
    \begin{eqnarray}
     |A_{\rm tot}|^2           & = &  |\sum_{i = 1}^{N} A_i \cdot \rm e^{2\rm i\pi \nu t_i}|^2 \\
                           & = &  \sum_{i = 1}^{N} |A_i|^2 \cdot | \rm e^{2 \rm i\pi \nu t_i}|^2 + \sum_{i = 1}^{N}\sum_{j = 1}^{N}A_i A_j \rm e^{2 \rm i\pi \nu t_i}\rm e^{2 \rm i\pi \nu t_j} .
\end{eqnarray}

Since the spots are considered incoherent, we have the following:

\begin{equation}
    |A_{\rm tot}|^2 \simeq \sum_{i = 1}^{N} |A_i|^2.
\end{equation}
Hence, $A_{\rm tot}$ corresponds to the sum of the photometric contrast contributions from the $N$ spots to the light curve. Using the results from Section\,\ref{sec:gaus_decription}, the height of the spectrum $\lnH$ is proportional to $|A_{\rm tot}\cdot B|^2$, which yields:
\begin{equation}
     \lnH \sim |A_{\rm tot}\cdot \life |^2 .
     \label{eq:lnH}
\end{equation}

Therefore, the amplitude of the power spectrum gives an indication of the coverage in terms of photometric attenuation (given by $A_{\rm tot}$ and due to the spot surface and spot temperature difference with its surroundings) during the entire observation time.
The dependency of $\lnH$ on $\life$  is the result of the contribution added by each spot transit over time.

When considering the presence of both single- and multiple-transit spots, their two contributions can be separated: $LC(t) = S_{\rm single}(t) + S_{\rm multi}(t)$.
Because of the linearity of the Fourier transform, these two contributions are also separated in the Fourier domain, but the three terms appear when considering their power spectrum. 

For the same reason as before, we can consider that the two contributions are incoherent and neglect the cross term: 

\begin{equation}
    |\widehat{LC}(\nu)|^2 = \widehat{S}_{\rm single}^2(\nu) + \widehat{S}_{\rm multi}^2(\nu).
    \label{eq:fitted_eq}
\end{equation}

\begin{figure*}
   \resizebox{\hsize}{!}
            {\includegraphics[]{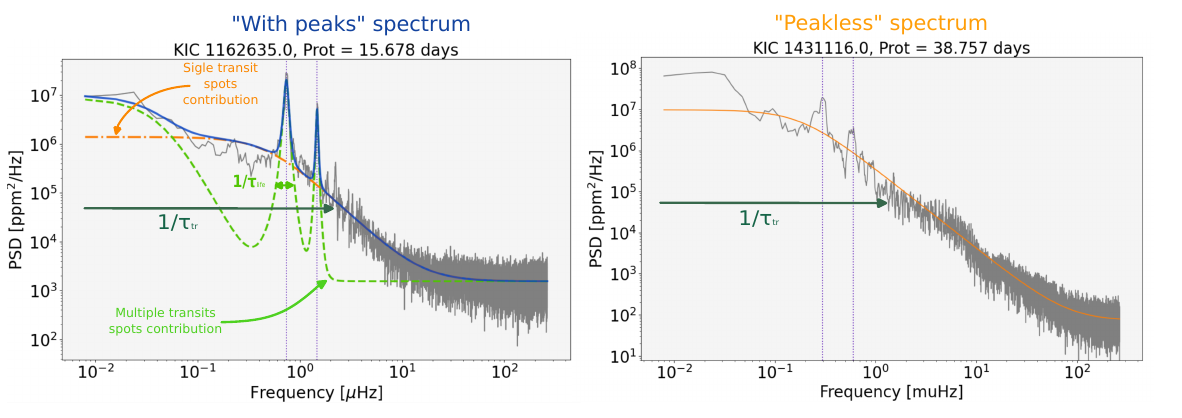}}
   \caption{
   The left panel shows an example of a power spectra density of a \textit{Kepler} LC (grey) “ with peaks” fitted by the model detailed in Sect.\ref{sec2}. 
   The transit component due to the single-transit spots is dash-dotted orange line, the multiple-transit spots component in a dashed light green line and the total fit in blue.
   The rotation peaks are marked by purple dashed lines. 
   The right panel shows an example of a peakless power spectra density of a star where the Harvey profile is used for the fit.}

   \label{fig:fit_example}
   \end{figure*}

There are hence two contributions to the PSD, as illustrated in the bottom of Fig.\,\ref{fig:model_ana} and Figure\,\ref{fig:fit_example} for a realistic general case: one due to the single-transit spots, which are dominated by the transit time ( orange box on Fig.\,\ref{fig:model_ana} and in dotted orange on Fig.\,\ref{fig:fit_example}), and another from the multiple-transit spots, which makes the rotation peaks appear and yields information on the spot mean lifetime (blue box on Fig.\,\ref{fig:model_ana} and in dotted light green on Fig.\,\ref{fig:fit_example}). The combination of these two contributions gives the LC fit in blue, which as we can see matches the large-scale trend of the data in grey.
Depending on the number of spots with a single or multiple transits, we can then identify two categories of spectra: spectra with clear rotation peaks (when spots with multiple transits dominate), hereafter the “with peaks” spectrum or without rotation peaks (when spots with a single transit dominate), hereafter the “ peakless” spectrum.
   
\subsection{Summary \correct{of the model}}
\label{sec:model_conclusion}

In summary, there are two profiles that have to be taken into account when a LC is fitted: first, the rotation peak profile $\widehat{S}_{\rm multi}^2(\nu)$ in light green in Fig.\,\ref{fig:fit_example}, which is due to the multiple-transit spots and is related to the spot lifetime $\life$. Second, the transit profile $\widehat{S}_{\rm single}^2(\nu)$ in light blue in Fig.\,\ref{fig:fit_example}, which is due to the single-transit spots and is closely related to  the transit time  $\ttransit$. In the case of a star covered only by spots that last less than one rotation period (single-transit-spot case), this profile is degenerated between the transit ($\ttransit$) and the lifetime ($\life$) components, and we recover a profile that can be better modelled by a Harvey profile described by only one characteristic timescale (see right panel of Fig.\,\ref{fig:fit_example}). 

Two types of spectra can therefore be inferred: those from stars covered mainly by multiple-transit spots where the two profiles are needed, and peakless ones where the two profiles are degenerated.  
This explains why some spectra show rotation peaks that are due to multiple-transit spots, while others do not. In the remainder of this paper, we will refer to 
“with peaks” spectra  (left panel of Fig. \ref{fig:fit_example})
and “peakless” spectra  (right panel of Fig. \ref{fig:fit_example}). 

Based on this total profile, three spot characteristics can be extracted from the spectrum of a LC: 

\begin{enumerate}
    \item The \textbf{transit proxy $\ttransit$}: the inverse of the width of the very low frequency component of the spectrum. It is the characteristic time of the transit and is thus close to $\prot/4$ but can vary depending on the rotation-axis inclination, limb-darkening, etc.
    \item The \textbf{lifetime proxy $\life$}: the inverse of the width at half-height of the rotation peaks, if they are visible. It provides information about the mean lifetime of the stellar spots during the period of observation.
    \item The \textbf{spot impact proxy $\lnH$}:  the height of the spectrum, the sum of $\lnH_{\rm single}$ and $\lnH_{ \rm multi}$. This proxy gives information about the total spot contrast coverage on the stellar disc during the observation time. 
    It is linked to two spots properties: their surface and their temperature difference compared to unspotted areas. This thus gives degenerate information about whether the spots are cold or small, or hot or large. This proxy is also sensitive to the inclination of the star. 
    However, given that we are looking at average trends, and that these two phenomena are linked to the magnetic field, this proxy allows us to make
    an estimate of the global spot activity of the star over the duration of the observation. 
    
\end{enumerate}

We emphasize that the two temporal proxies ($\life$ and $\ttransit$) are only relevant for the stars showing rotational peaks, while they are degenerated for the stars not showing peaks.

%
\section{Validation of the model with simulations}
\label{sec:simu}

In this section, the model described in Sect.\ref{sec2} (Eq.\,\ref{eq:fitted_eq}) is validated against simulated LCs based on the analytical model of \citet{Dorren97}, which describes how a single circular spot impacts a LC as a function of different stellar and spot properties.
By computing the impact of several spots and summing these contributions, this model permits us to simulate LCs. We can modify several parameters such as the number of spots, their lifetime, their distribution on the stellar disc, their radii, the rotation period of the star, the possible differential rotation (DR) and the inclination of the star rotation axis. 
The code also uses the Gaussian function given by Eq.\,\ref{eq:mosser_description} \citep{Mosser2009} to describe the intrinsic evolution of the spot. 
The parameters explored are: the influence of the rotation period ($\prot$), the spot lifetimes, their number, and  their radius. 
The spectrum of the simulated LCs is then computed and fitted using the Maximum Likelihood Estimator (hereafter MLE) described by \citet{Toutain1994}. 

\subsection{Method}
\label{sec:algo_description}
We compute several sets of LCs for various stellar parameters. Each set is composed of an ensemble of 200 simulated LCs, using the same stellar parameters but a different random set of spots on the stellar disc in terms of time and location (following a normal distribution).

For the spot lifetime, we performed three different types of simulations: one with only spots that last less than the rotation period to produce peakless spectra ($\life = 1/6 \prot$ hereafter ''single transit''). It corresponds to the single-transit-spot case and is close to solar observations. 
Another type with only multiple-transit spots (i.e. $\life = \prot$, hereafter ''multi-transits''), where the spots have more than 5 visible transits with a significant depth in the LC (since the lifetime and transit time were defined as the width at half height). 
The last type has lifetimes that randomly vary according to a uniform law between $1/6 \prot$ and $\prot$ (hereafter 'mixed'). 
The simulations with only multiple-transit spots are not  realistic, since there should be shorter additional photometric contributions coming from physical phenomena other than spots (for instance,  faculae or flares), but they serve as an extreme test case.
These simulations were also carried out for 3 rotation periods of 10, 20 and 30 days.
For a given set, all the spots have the same area and the LCs have a duration of two years, as for most of the \textit{Kepler} LCs that will be used subsequently.

Since the contrast of the spot is degenerated between temperature and area, we choose to only change the spot surface area to study how $\lnH$ varies in our set of simulations.
Other sets have been generated, where the size and number of spots varied, as well as the inclination of the stars. 
For each normalised LC, a Gaussian noise with a deviation of $\sigma = 10^{-5}$ has been added. 
This yields to 23 sets of simulations, each set comprising 200 simulated LCs.
A summary of the parameters associated with each set is given in Table \ref{table:simulation}.

\begin{table*}
\caption{The simulation parameters for the validation of the model. Each line corresponds to a different set with its associated number (each set comprising 200 LCs with different random generation of spots location over time). Each set can differ from the others based on the rotation period of the star, the total number of spots, their typical lifetime (single transit, multiple transits, or mixed), their typical radius, or the inclination of the star. For the mixed LCs, we provide an interval of lifetimes rather than a single value.}  
\label{table:simulation}      
\centering          
\begin{tabular}{ccccccc} 
\hline\hline               
N° &  $\prot$ (days) & Number of spots &  Lifetime ($\prot$)&  Spot radius (deg)& Inclination of the star (deg) \\
\hline     
          
          0& 20 & 35 &  1/6        & 1 &  90\\ 
          1& 20 & 35 &  1          & 1 &  90\\ 
          2& 30 & 35 &  1/6        & 1 &  90\\ 
          3& 10 & 35 &  1/6        & 1 &  90\\ 
          4& 20 & 70 &  1/6        & 1 &  90\\ 
          5& 20 & 35 &  [1/6,1]    & 1 &  90\\ 
          6& 10 & 35 &  1          & 1 &  90\\ 
          7& 30 & 35 &  1          & 1 &  90\\ 
          8& 30 & 35 &  [1/6,1]    & 1 &  90\\ 
          9& 10 & 35 &  [1/6,1]    & 1 &  90\\ 
         10& 10 & 70 &  1/6        & 1 &  90\\ 
         11& 20 & 20 &  1/6        & 1 &  90\\
         12& 20 & 70 &  1          & 1 &  90\\
         13& 20 & 20 &  1          & 1 &  90\\
         14& 20 & 50 &  1/6        & 1 &  90\\
         16& 20 & 35 &  1/6        & 2 &  90\\
         17& 20 & 35 &  1          & 2 &  90\\
         18& 20 & 35 &  [1/6,1]    & 2 &  90\\
         19& 20 & 30 &  [1/6,1]    & 10 & 90\\
         20& 20 & 30 &  [1/6,1]    & 1 & 60\\
         21& 20 & 30 &  [1/6,1]    & 1 & 30\\
         22& 20 & 30 &  [1/6,1]    & 1 & 0\\
         \hline
    \end{tabular}
\end{table*}

Once generated with its associated stellar parameters, each LC spectrum has been fitted with the model proposed in this paper, i.e. a combination of a function for the transit profile (which represents the dominant contribution of the single-transit spots), plus a function for the rotation peaks (which represents the dominant contribution of spots with multiple transits, see Eq. \ref{eq:fitted_eq}). 

For these simulations, each spectral component associated with the transit and the time evolution  profiles (i.e. each rotation peaks) can be expressed by the following Gaussian profile:
\begin{equation}
    G(t) = A \rm e^ { -  (\frac{(\nu - \nu_0)\pi\tau}{\ln({2}}  )^2}, 
    \label{aq:gaussian}
\end{equation}
since the temporal evolution profile of the spot is a Gaussian, and their transit can also be approximated by a Gaussian. 
The set-up of the simulations is thus close to the description made in Sect.\ref{sec:gaus_decription}, but here we only fit the contribution of the multiple-transit spots $\widehat{S}_{\rm multi}(\nu)$ (i.e. the rotation peaks) by using each time the Gaussian form of Eq. \ref{aq:gaussian}. We do not take into account the transit component in $\widehat{S}_{\rm multi}(\nu)$ since it only appears in the amplitude of the peaks, which is a free parameter in the algorithm to help it converge. 
We choose to fit 2 rotation peaks plus the contribution centred at the zero frequency, because in most of the cases the higher harmonics are not clearly visible above the noise.
For $\widehat{S}_{\rm single}(\nu)$, there is only one Gaussian that describes the transit time, since there are no clear rotation peaks.
Moreover, the simulated LCs have also been fitted by the pseudo-Lorentzian model proposed in Eq.\,\ref{eq:Harvey} of \citet{Harvey93} in order to compare our new model with this already existing one.

\subsection{Results}

\subsubsection{Spot impact proxy}

The sets of simulations used here to test Eq. \ref{eq:lnH} are listed in Table \ref{table:simulation}. 

   \begin{figure}
   \centering
   \includegraphics[width=\hsize]{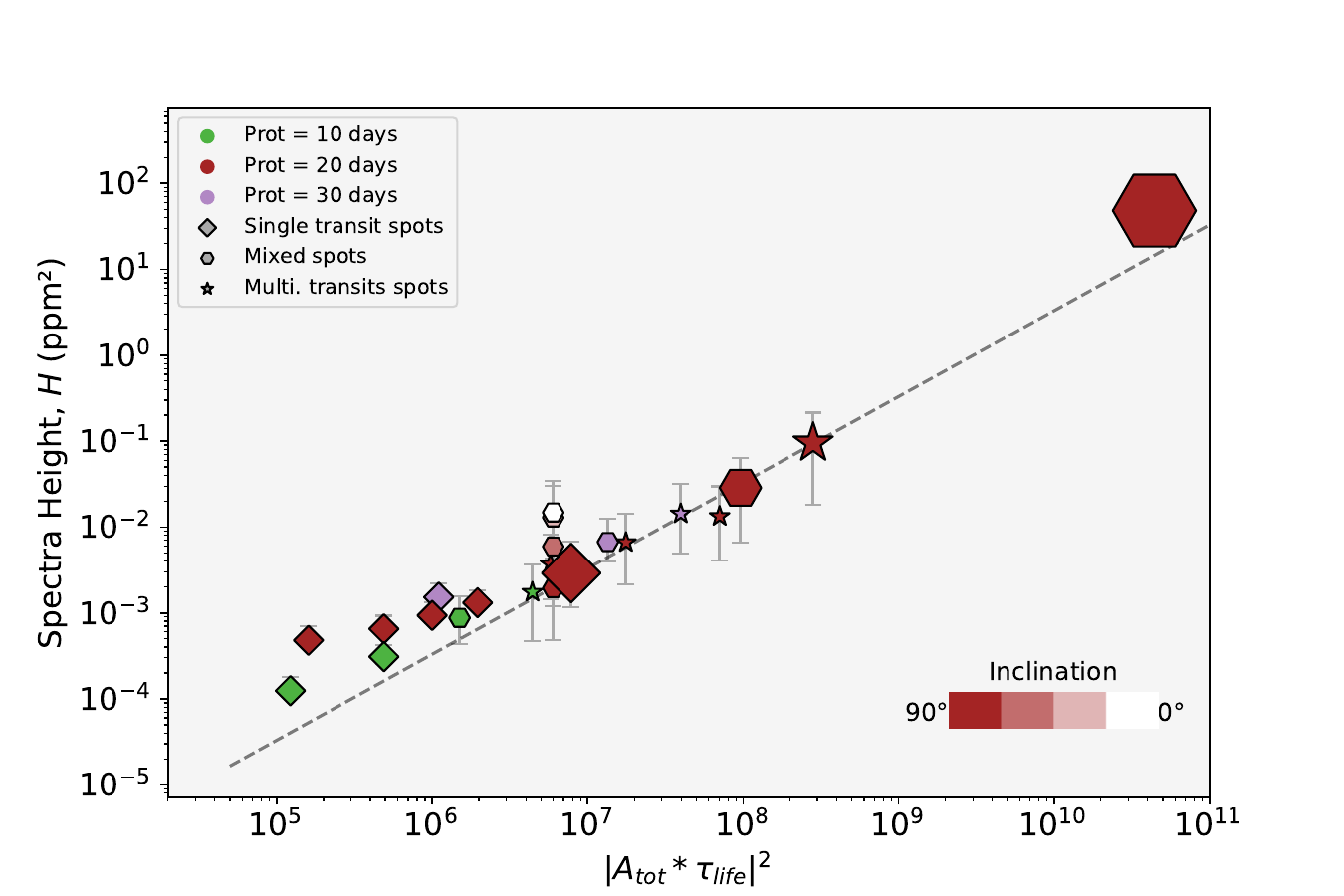}
      \caption{Median height of the spectrum ($\lnH$) of each set of simulations as a function of $|A_{\rm tot} \life|^2$ (see Eq. \ref{eq:lnH}). 
      The colour of the dots gives the rotation period of the star (10 days in green, 20 days in red and 30 days in purple), while the shape gives the type of spot lifetime (diamonds for the peakless stars, hexagons for mixed spots and 5-pointed stars for stars with rotation peaks spectra). The size is proportional to the surface of the spots. The large \correct{hexagon} on the top-right corresponds to set 19 with spots radii of 10 degrees.
      The error bars are based on the first and third quartile. 
      The points with a shading towards white are the results for the 3 simulations where the inclination varies (sets 20, 21 and 22), white being for 0° inclination (set 22).}
         \label{fig:simu_act_proxy}
   \end{figure}

Fig.\,\ref{fig:simu_act_proxy} shows a clear correlation for all sets between the median estimate of $\lnH$ and the surface covered by the spots.
The size of the individual spots and the temporal behaviour of spots (mixed and multi-transit cases) do not impact significantly the trend. 
However, simulations with only single-transit spots show a different dependency, especially when considering different rotation periods, where $\lnH$ tends to be overestimated. This may be due to the fact that, in the case of the peakless spectrum, the model does not lift the degeneracy between the signature of the lifetime and the transit, and the approximations  made to express $\lnH$ in Eq. \ref{eq:lnH} are less valid. The impact proxy $\lnH$ is still an increasing function of $|A_{\rm tot}\cdot \life |^2$ for a given inclination, but nevertheless seems to slightly deviate from a linear relation. 
\begin{figure*}
   \resizebox{\hsize}{!}
            {\includegraphics[width=\hsize]{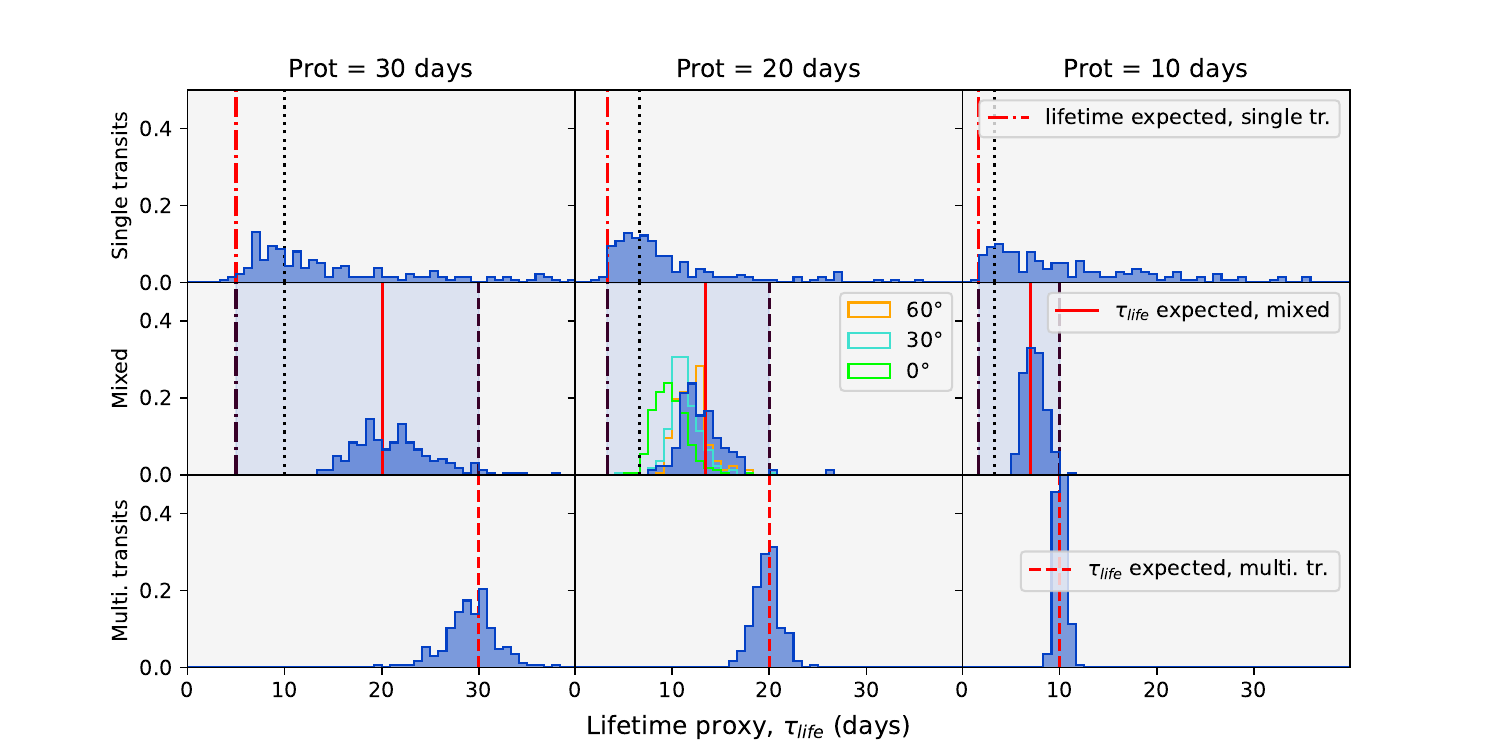}}
        
      \caption{ Normalised histograms of $\life$. Each column corresponds to a different rotation period, and each row to a type of lifetime (single-transit, multi-transit and mixed spots). Here, only the results for the new model are shown, since it cannot be retrieved using the Harvey model.  
      For each panel, \correct{the red lines show the expected value.} The vertical black dash-dotted line represents the lifetime \correct{value} for the single-transit spots and the dashed \correct{the  one for the spots that last several transits (i.e. $\life = 1 \prot$)}.
      In the case of mixed spots, \correct{these two lines are represented}, as they show the range of possible values. This space has been coloured in \correct{blue} for clarity. The full line represent the mean value of the lifetimes.
      For single-transit and mixed-spots simulations, the vertical black dotted line represents the value of lifetime where two transits are visible in the LC. 
      The stepped histograms in \correct{green, blue and orange} correspond to the simulations where the inclination of the rotation axis varies (respectively to 0, 30 and 60 degrees). 
      }
         \label{fig:tau_peaks_hist}
   \end{figure*} 
Finally, we have tested the impact of the inclination of the stellar rotation axis using three more sets of simulations with mixed spots (sets 20, 21 and 22) for inclinations of 60° (light red), 30° and 0° (white)  compared to the line of sight. 
The estimations of $\lnH$ for an inclination of 30° and 0° are overlaid. $\lnH$ is higher compared to the case with a 90° inclination. We observe that the closer the axis is to 0°, the higher is the value of $\lnH$.  
This deviation from Eq.\,\ref{eq:lnH} is due to the approximation $\ttransit \sim \prot / 4 $ made in the expression of $B$ in Eq.\,\ref{eq:gaus_approx}, which is no longer true for inclination angles near to 0° where $\ttransit$ is closer to $\prot$. 
The spot impact is therefore higher in these configurations, since the mean transit time is longer as the mean of the spot signal over time is higher.

\subsubsection{Lifetime proxy}

The sets used in this section are 0, 1, 2, 3, 5, 6, 7, 8, 9, 20, 21 and 2.
The results for this proxy are represented in Fig.\,\ref{fig:tau_peaks_hist}.
First, we can see that the rotation period does not affect the spot lifetime measurement much (similar results between the different columns). However, for mixed and multiple-transit spots (middle and bottom rows), the distribution is as when the rotation increases. 

Concerning the multiple-transit-spot cases (bottom row), the mean lifetime is as expected equal to the rotation period (see Table \ref{table:simulation}), with a deviation of approximately 10 $\%$.

For the mixed spots cases (middle row), the distribution is spread between the maximum value and $\prot/3$, This value corresponds to the case where the spots have a two visible transits ($\life = 2 \cdot 1/6 \prot$) in the LC and where the rotation peaks begin to appear.
The median value is therefore located at the middle of this interval ($2/3 \cdot \prot$), the value around wihch the distribution is located .

In the case of single-transit-spot simulations (top row), the fit has a weaker agreement with the proxy. Since in this case there are none or very weak rotational peaks, as in the case of the transit proxy, the fitting algorithm encounters more difficulties to converge properly.
However, there is still an accumulation around $\prot$/3 (dotted line in Fig.\, \ref{fig:tau_peaks_hist}). 
Here also, this is because some single-transit spots can sometimes produce two transits in the LC, typically when they appear right before passing behind the limb, and then re-appear at the opposite limb and fade very quickly. This induces a periodic signal that can be seen in the spectra, even for a single-transit spot.
Despite this, the results show that the lifetime values found by the model are spread over a wide range of values, which also covers the lifetimes of multiple-transit-spot cases. In the case of peakless spectra produced by single-transit spots, the model therefore fails to converge reliably to a correct lifetime value, which is expected: in this case, both the lifetimes and transit times of the spots are degenerate.

The simulations where the inclination of the stars have been varied are also presented in Fig. \ref{fig:tau_peaks_hist}. 
We can see that for the minimum inclination angle (0°), the values found for $\life$ are biased towards shorter lifetimes, by a relative value of 30$\%$. For the other two inclinations, the values found remain very close to the values without inclination.

\subsubsection{Transit proxy, $\ttransit$}
\begin{figure*}
   \resizebox{\hsize}{!}
            {\includegraphics[]{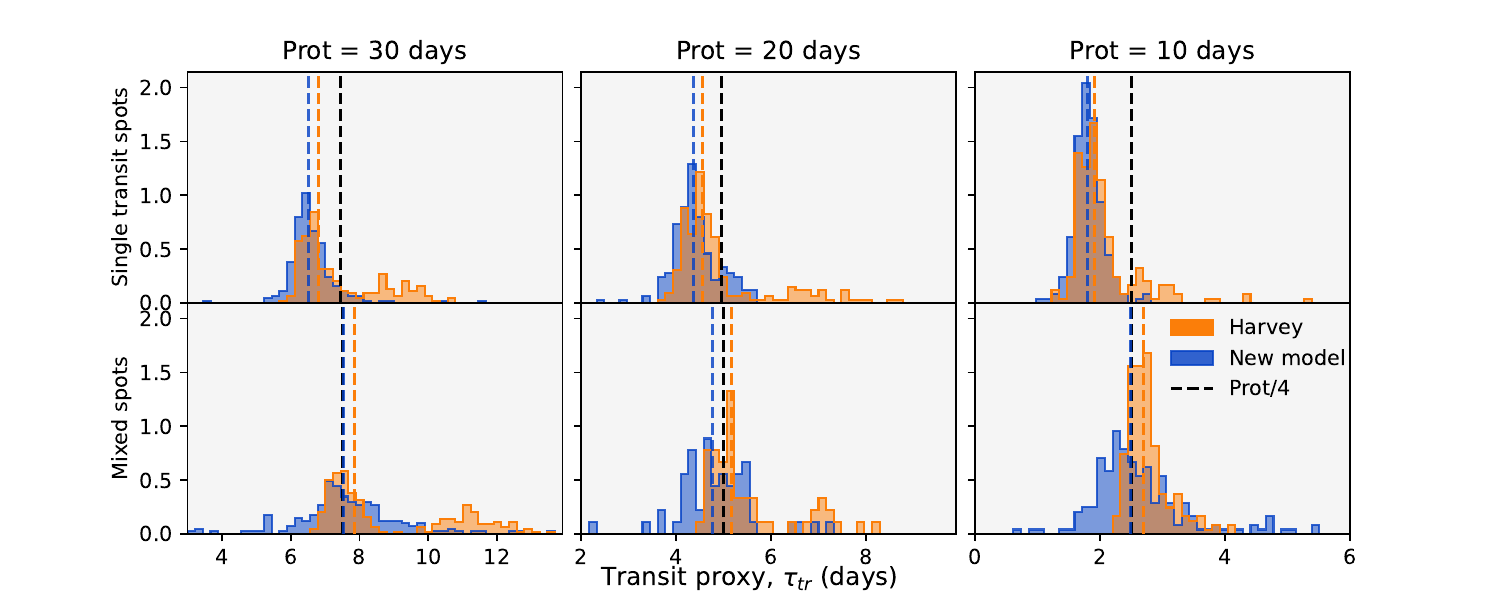}}
      \caption{ Same as Figure \ref{fig:tau_peaks_hist}, but for the transit proxy $\ttransit$.  The histograms in blue correspond to the value found by our model, and the ones in orange to the value found by the model of Harvey. The vertical black dashed lines indicate the expected value for each set of simulations, which corresponds to $\prot/4$. The coloured dashed lines represent the median value of each distibution.
      }
         \label{fig:tau_hist}
   \end{figure*} 
In this section, only the sets with varying lifetimes and rotation periods are used, since $\ttransit$ is only sensitive to temporal parameters (see 0, 1, 2, 3, 5, 6, 7, 8, 9). 
For this proxy, we compare the results from our new model proposed in this paper to those from \citet{Harvey85}. Indeed, the width of the pseudo-Lorentzian profile has been related to the characteristic time of the magnetic activity; but according to our model, this value is closer to the transit time.  
Fig. \ref{fig:tau_hist} shows the $\ttransit$  distributions for each set of simulations. 

For single-transit-spot cases (top rows), both models underestimate $\ttransit$.
This discrepancy can be explained by the fact that, in this case, the transit time and the lifetime are degenerate. Although the rotation peaks do not appear clearly, they still contribute to the broadening of the spectrum and their presence is included in the fit function. 
Since the contribution of the transit is not distinguishable from the rotation peaks, the fit converges to a biased value.

For the mixed-spots case (bottom rows), the expected value is found by the algorithm in the three  different rotation cases, but with shifted values for the Harvey model for fast rotators. 
The Harvey model also has an accumulation of points for higher values, in both single-transit-spot and mixed cases. 
This is due to the fitting algorithm, which in some cases will converge to the fundamental peak, which is located at a shorter interval than the transit value, because these are not taken into account in the Harvey model.

The case with only multiple-transit spots is not shown here because the model we use cannot recover a correct transit time without the contribution of the single-transit spots that we have chosen not to add in these simulated LC.  
However, according to our analytical description, the transit time is still present when following an envelope passing through the maximum height of the rotational peaks. This particular case is discussed in Appendix \ref{appendix:longspot_simu}. 

This result shows that the Harvey profile is closely related to the transit time of the spots (and hence, to the rotation period), and not to the characteristic time of the activity, or more specifically the spot lifetime.
The results obtained by \citet{Aigrain2004} can be explained by the fact that, in the case of the Sun, the spot lifetime is close to the rotation period. It is a “peakless spectra” case, where the rotation peaks are less apparent and the dominant profile is the one from the transit.

The inclination of the stellar rotation axis also modifies the estimation of $\ttransit$. The distribution of values is significantly wider and shifted towards higher values, as expected (the transits are longer when the star is inclined). The results are not included on Fig. \ref{fig:tau_hist}, but the median values of the distribution are shifted to 6.2, 15 and 20 days for an inclination of 60°, 30° and 0°, respectively.  

\subsubsection{Summary of the simulations results}
\label{sec:conclusionSimu}

Analysing the profiles of the spectra in the Fourier domain allows us to validate the interpretation of the proxies proposed in Sect.\ref{sec:model_conclusion} and determine their limitation :

\begin{itemize}
    \item The spot impact proxy $\lnH$ shows a good correlation with the contrast induced by the surface covered by the spots. The inclination of the rotation axis can have an impact on this proxy since the spots appear longer, but we can nevertheless assume that the catalogues of observed stars only contain a small fraction of highly-inclined stars, as mentioned before and according to \cite{2023Bowler}.
    \item The lifetime proxy $\life$ is correlated with the mean value of the spot lifetime, when more than two transits per spot are observed \correct{, i.e. when $\life > \frac{1}{3} \prot$ according to the values of simulation.} Otherwise, the model has difficulty to converge to an exact value, since $\ttransit$ and $\life$ are degenerated. It is therefore only possible to evaluate this proxy if there are significant rotation peaks in the spectrum under study. \correct{This means that the estimate of the mean lifetime is biased by the rotation rate: for fast rotators, we will be
biased towards shorter $\life$ (if any measured) since $\life$ cannot be estimated if longer than 1/3$\prot$.} 
    If there is no \correct{significant rotation peaks}, the Harvey model is sufficient, but the calculated characteristic time cannot be used to estimate an average spot lifetime.  
    
    The inclination of the rotation axis can also underestimate $\life$ when the angle of inclination is very small.
    \item The transit proxy $\ttransit$ is well correlated with the mean transit time of spots. It is underestimated for LCs with only single-transit spots since there is a degeneracy between the lifetime and the transit time in this case. For stars with multiple-transit spots, its value can be recovered through the contribution of single-transit spots to the spectrum, but also in the height of the rotational peaks.
    However, to simplify the fitting algorithm, each peak has been adjusted individually so this component has not been included in the fit function. 
    The inclination of the stars may bias the estimation of $\ttransit$ if this inclination is very pronounced.
    However this proxy will not be a primary focus of this paper.

\end{itemize}

For the remainder of the analysis using \textit{Kepler} data, it is thus necessary to differentiate between spectra with and without observable peaks. This distinction ensures that we apply the model only when it can estimate lifetimes effectively.
In the case of “peakless” spectra, we will apply the Harvey model, which still provides an estimate of $\lnH$.

\section{Application to the \textit{Kepler} data}
\label{sec:result} 

\subsection{Datasets}
\label{subsec:kepler_data}
The LCs used in this work are part of the 150 000 stars observed by the \textit{Kepler} mission. 
They have been taken from the MAST (Mikulski Archive for Space Telescopes) data archive. 
Although the \textit{Kepler} mission made observations over a 4-year period, the light curves used here last for an average of two years, due to the data reduction employed.  
These data have been first detrended by the Presearch Data Conditioning-Maximum A Posteriori \cite[PDC-MAP, e.g.][]{Jenkings2010, Smith2012, Stumpe2012}. 
Another treatment has been applied, described by \citet{Peralta2018}\footnote{\url{https://ssi.lesia.obspm.fr/sites/ssi/IMG/pdf/user_guide_v1-1.pdf}}, where the main goal was to fill the gap between the quarters using a simple stitching method, to remove the outliers and to apply a linear trend correction.

Two catalogues of data have been used in this work. The first is from \citet{McQ2014}, which provides an estimate of the rotation period for 33\,040 \textit{Kepler} stars using the autocorrelation function method \cite[ACF, method detailed by][]{McQ2013}. 
The catalogue also contains other stellar parameters such as the effective temperature $ \rm T_{ \rm eff}$, the $\log g$ extracted from the \textit{Kepler} Input catalogue (KIC) or estimated from \citet{Courtenay2013},
and an estimation of the mass, by using the isochrone from \citet{Baraffe1998}, which was computed specially for low-mass stars. 
The sample is composed only of main-sequence stars, where the known eclipsing binaries and stars with a planet candidate (KOI, \textit{Kepler} Objects of Interest) were removed. 

The second catalogue used here is the concatenation from \citet{Santos2019} and \citet{Santos2021}. 
This sample is made of 55\,232 \textit{Kepler} main-sequence stars and subgiants from spectral type M to F. For this paper, we removed the subgiants (5544 stars) from this sample to focus only on the main-sequence stars. 
This catalogue also provides an estimation of the rotation period by using a combination of three methods: the first one is the ACF, the second one is a global wavelet power spectrum analysis and the third one is based on a composite spectrum, which is a combination of the two first ones \cite[see][]{Cellier2017}. 
The catalogue also contains an estimate of the mass, $\rm T_{\rm eff}$ and $\log g$ from \citet{Mathur2017}. 
Most of McQuillan's stars can also be found in the Santos set. A total of 25 855 new stars are added from this catalogue that are absent from McQuillan's catalogue.

It is important to point out that the methodologies used by these two studies to measure rotation periods can induce errors in their estimates up to a factor of two. Despite the meticulous efforts in these papers to mitigate this issue, techniques such as autocorrelation function (ACF) analysis or peak identification can occasionally converge to a harmonic rather than the actual period. This error can also come from the stellar configuration where two large spots are present on the star at opposite locations. Some periods used in this article may therefore be incorrectly estimated.

The difference between these two catalogues relies on the method used to determine $\prot$. Since \citet{McQ2014} used the ACF only, one can assume that the procedure of \citet{Santos2019} and \citet{Santos2021} is more sensitive to the variability and includes fewer active stars. 
For these two catalogues, we also used the Gaia mass estimation by FLAME\footnote{\url{https://gea.esac.esa.int/archive/documentation/GDR3/Data_analysis/chap_cu8par/sec_cu8par_apsis/ssec_cu8par_apsis_flame.html}}, when available.
However, since we observed a certainly biased accumulation of stars with mass around 0.7\,$\rm M_{\odot}$ in the \textit{Gaia} data, we chose to only use \textit{Gaia} masses higher than 0.8\,$\rm M_{\odot}$ and kept the mass value from the two catalogues otherwise. 

In the main part of this section, we will analyse the results simultaneously for both McQuillan's and Santos' datasets.

\subsection{Algorithms for the \textit{Kepler} data }

Two algorithms were developed to analyse the Kepler data. The first is a sorting algorithm that classifies spectra with and without peaks by examining the first four rotation peaks using three different methods.
The first method analyses the derivative of the spectrum to detect significant changes in sign.
The second method checks whether the peak heights are at least three times greater than the overall standard deviation of the spectrum.
The third method evaluates whether each peak height is at least three times higher than the local standard deviation immediately after the peak.
Thus, 12 tests (3 methods applied to 4 peaks) are performed on each spectrum.
A spectrum is considered to have peaks if all three methods agree on at least two of the first three peaks, or if 8 out of the 12 tests are positive. Conversely, a spectrum is classified as lacking peaks if none of the methods converge on any of the peaks, or if fewer than 5 tests are positive.
This algorithm was tested on a sample of 1\,000 spectra that had previously been sorted by visual inspection. 
In 87\,$\%$ of cases, the algorithm found the same results as visual sorting. 

The second algorithm developed was the fitting algorithm.
The same algorithm as before (Sec.\,\ref{sec:algo_description}) has been used to fit the LC power spectrum, but this time a pseudo-Lorentzian (Eq.\,\ref{eq:Harvey}) profile has been used for $S_{\rm single}$ and a simple Lorentzian ($\alpha= 2$) for the rotation peaks, i.e. $S_{\rm multi}$ in the case of the stars identified as “with peaks” by the sorting algorithm. These functions indeed allow more degrees of freedom with the $\alpha$ exponent, and were thus more suited for actual observations.
Since the aim is to retrieve a width at half-height and compare global trends for the data set, and not to find exact lifetime values, changing the fit function should not impact the results, as explained in Sect.\ref{sec:gaus_decription}.  
The fitted spectrum takes into account two rotation peaks because that is the most common case observed. 
However, in some cases, the LC spectrum can show more than two peaks that can bias the fit. 
Given that, it will especially impact the estimation of the transit time (see Sect.\ref{sec:simu}) that is not the main proxy of interest here, this compromise can thus be considered as acceptable. 
In the fitting process, the rotation period is fixed to the value extracted from the catalogues. 
The Harvey profile (i.e. only a pseudo-Lorentzian) has been used for the stars with peakless spectra.
We recall that an example of these two fits is shown in Fig.\,\ref{fig:fit_example}.

The total catalog that we were initially using consisted of more than 50\,000 stars. 
After running the sorting algorithm, 57.4\,$\%$ of the data were classified as spectra with peaks, 37.1\,$\%$ without and 5.5\,$\%$ as uncertain for the whole dataset. We did not use the spectra classified as “uncertain” in our study which did not impact our sample since they represent a very small portion of the sample.
After running the fitting algorithm on the data, we removed the spurious fit values (zero values, infinite values, negatives, etc.). \correct{We also removed the stars that do not respect  the  predicted criterion for the appearance of rotation peaks (see Sect.\ref{sec:conclusionSimu}). In fact, we observe a small group of values below $\life / \prot \approx 0.35$, which, after a visual verification, corresponded to peakless spectra that the sorting algorithm did not classify properly.} We end up with a sample of 33\,755 stars.
More technical details about this selection are given in Appendix \ref{appendix:technical_algo}.



 \begin{figure}
   \centering
   \includegraphics[width=0.96\hsize]{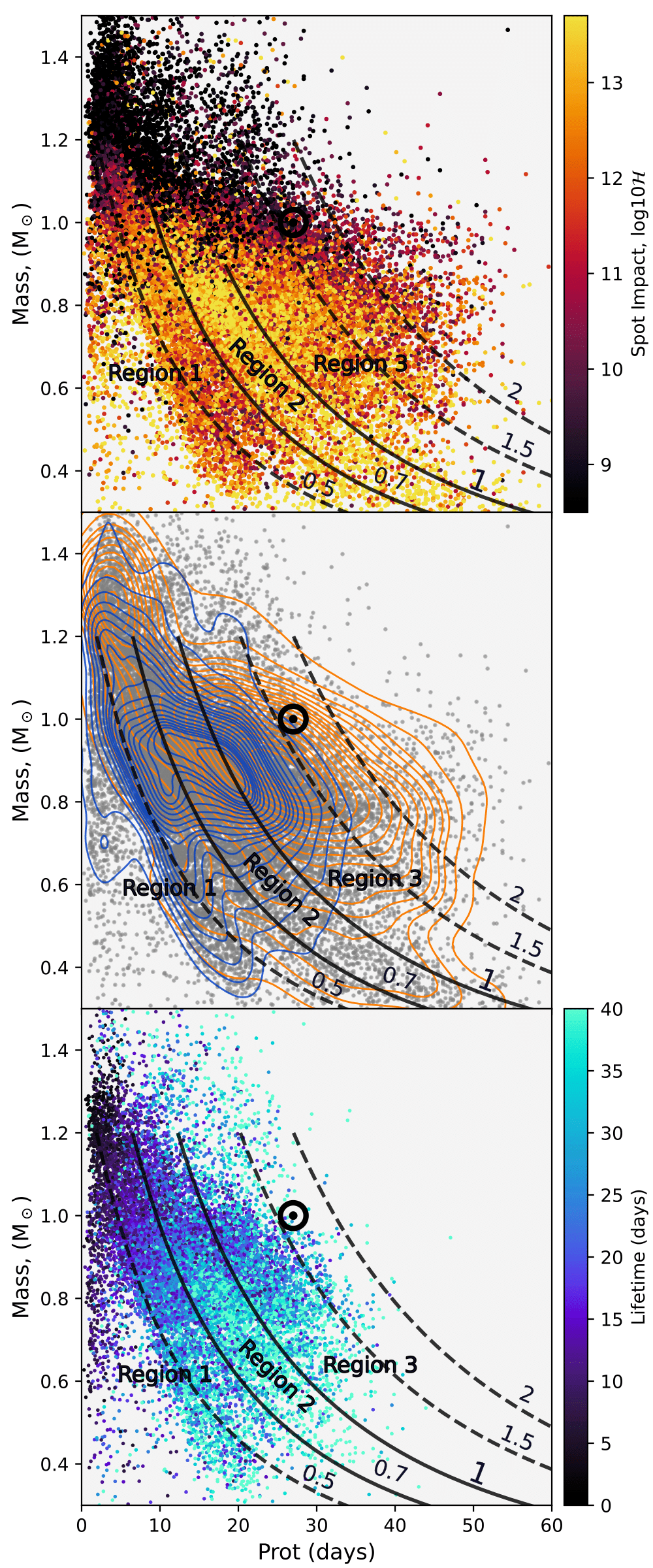}
      \caption{Top panel: spot-impact proxy for the stellar sample as a function of mass and rotation period. The colour scale shows the impact proxy $\lnH$ with a logarithmic scale: yellow is for a large spot impact (active stars), black for low spot impact (less active stars). 
      The black lines represent the iso-lines of the Rossby number.
      The $\odot$ point indicates the location of the Sun in this diagram.  
      Middle panel: same figure with contour lines showing the locations of the stars with “peakless” or “with peaks” spectra, respectively in orange and blue. The grey dots represent the whole sample.
      Bottom panel: same as the top panel, but with the lifetime proxy $\life$. Only stars with spectra with peaks are presented.}
         \label{fig:lnh_mck}
   \end{figure}

\subsection{Results}\label{sc:results_kp}
In order to present the results obtained for $\lnH$ and $\life$ and present a landscape of magnetic spot activity as a function of different stellar parameters, we have chosen to start by plotting them as functions of stellar rotation and mass (Fig.\,\ref{fig:lnh_mck}).   
The position of groups with and without peaks is also shown. 
These distributions are compared to the stellar Rossby number:   
\begin{equation}
   \rm Ro = \frac{\rm \prot}{\tau_{\rm c}}   ,
\end{equation}
where $\tau_{\rm c}$ is the turn-over timescale for the convection. This number helps determine which effect is dominant between the rotation and the turbulent convective flows, which are the main two processes involved in the dynamo mechanisms responsible for the generation of the magnetic activity.
This number is computed using the mass-dependent relation of \citet{Wright2011}, where  $\tau_{\rm tc}$ has been computed using the colour-conversion relation of \cite{Pecaut2012}. 
This relation has been established only for stars with masses between $0.2$ and $1.2 \, \rm M_\odot$ and is no more valid for higher masses in the sample.   
We begin by describing the results for $\lnH$, then $\life$, and finish with a summary of the different trends that we observe as a function of the stellar mass and the Rossby number.

Regarding the estimation of the transit time and the characteristic time associated with the Harvey profile, we refer the interested reader to Appendix C as the physical information provided by these parameters is not essential to the following discussion.

\subsubsection{Spot impact proxy $\lnH$}
\label{subsec:spot_impact_result}

The top panel of Fig.\,\ref{fig:lnh_mck} shows the distribution of $\lnH$ values.
We recall that we are using the estimations of our model for spectra with peaks and the model of Harvey for peakless spectra.

Different trends emerge from this Fig.. First of all, there is a clear distinction between stars below and above one solar mass. For stars above 1 $\rm M_\odot$, the spot impact is much lower. 
For the less massive stars, the spot impact evolves differently depending on the Rossby number. The most active stars (in yellow) are seen in a banana shape between the iso-Rossby lines $\rm Ro= \, 0.7$ and $1$ (hereafter Region 2). We will refer to this area as the high-activity regime. 
Then, on each side of the high-activity regime, there are two zones where the spot impact proxy shows intermediate values (orange). 
The one on the faster rotator side (left) is located at smaller Rossby number values of $\rm Ro =$\,0.5 (hereafter Region 1). 
The other one, on the longer period side (right), corresponds to Rossby numbers greater than 1 (hereafter Region 3).

\subsubsection{Localization of peakless and with peaks spectra}

The middle panel of Fig. \ref{fig:lnh_mck} presents the location of stars with peakless spectra compared to those with peaks. 
The “with peaks” population (blue) concerns mainly fast rotators and stars below 1\,$\rm M_\odot$. They are massively present around the isoline of $\rm Ro = 1$ and below, as well as on the branch of the fast rotators described by \cite{McQ2014}, who observed that the rotation period distribution is divided into two branches for low effective temperature stars.
The stars with peakless spectra (orange) are either slow rotators located above the isoline $\rm Ro = 1$ or more massive stars ($M > 1 \rm M_\odot$) that have small $\prot$.

\subsubsection{Lifetime proxy $\life$}
\label{subsec:lifetime_mck}

The bottom panel of Fig. \ref{fig:lnh_mck} presents the results for the mean lifetime of stars with rotation peaks.
Similar to the spot impact, this proxy is related to the stellar mass: the more massive stars have very short lifetimes for their spots, while the low mass stars have longer ones.

\begin{figure}
   \centering
   \includegraphics[width=\hsize]{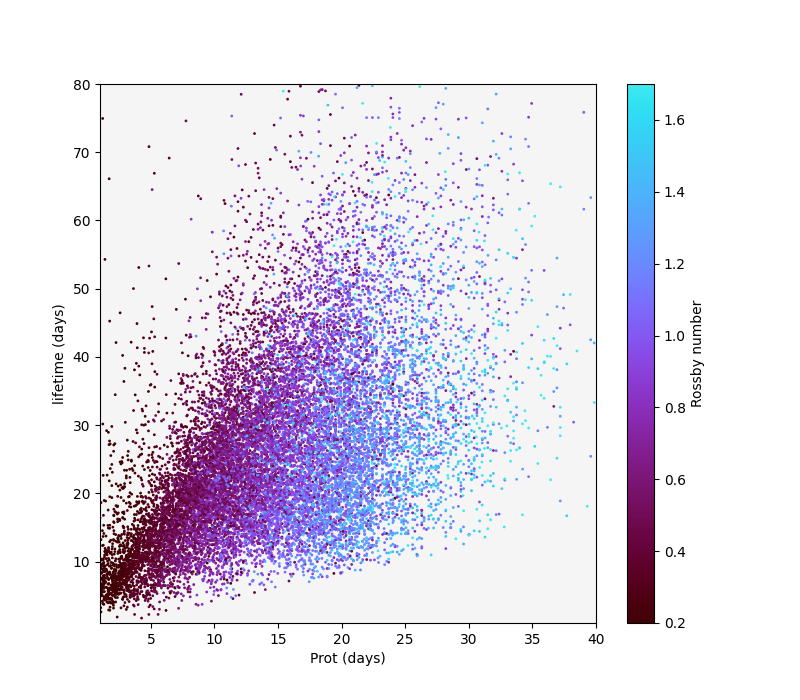}
      \caption{The estimated mean lifetime $\life$ as a function of the rotation period for stars with spectra with rotation peaks. 
      }
         \label{fig:life_prot}
   \end{figure}

There is also a dependence on the rotation period: slow rotators have longer lifetimes than fast rotators, \correct{which is not surprising, given the Sect. \,\ref{sec:conclusionSimu} conclusions:  the model struggles to reliably estimate $\life$ for values below $\prot /3$. Consequently, the slower a star rotates, the more challenging it becomes to accurately determine shorter lifetimes}. 
This relationship between rotation and lifetime is illustrated more clearly on Fig. \ref{fig:life_prot}. 
The same trends has been observed by \cite{2022Basri} for the MacQuillan stars that we have in common, but we do not retrieve the diffuse second branch that they described. 
Similarly to the spot impact proxy, but less distinctly, $\life$ of the less massive stars is organized according to the Rossby number: longer lifetimes appear between $\rm Ro = 0.7$ and $\rm Ro = 1$, in the same location Region 2 described in Sect.\ref{subsec:spot_impact_result} where the higher $\lnH$ are found. 
In Region 1 (below $\rm Ro = 0.7$), longer lifetimes can also be seen under $0.6 \, \rm M_\odot$.
This can also be seen on Fig. \ref{fig:life_prot}.

But what can we say about the “peakless” group of our sample? The middle panel of Fig. \ref{fig:lnh_mck} shows that for the less massive stars (M $< 0.6 \rm M_\odot)$, they have mainly slow rotation period. 
For these particular stars of our sample, we do not have  estimations of $\life$ but we still know that these stars have spot with lifetimes shorter than or equal to their rotation periods (see Sect.\ref{sec:model_conclusion}). 
As the definition of lifetime is the width at half-height, this means that the mean lifetime of spots on these stars would be shorter than their rotation. 
This means that this population located at large Rossby values (i.e. above $\rm Ro \sim 1.2$) has shorter lifetimes than the majority of those observed for the group with rotation peaks in this area.
This statement is not so obvious for high-mass stars as there are very few stars in the group with peaks for comparison. However, we can intuitively expect that these \correct{starspots} likely have short lifetimes, given their rapid rotation.

\subsection{Summary: activity and Rossby number}
\label{subsec:summary_result}

\begin{figure}
   \centering
   \includegraphics[width=\hsize]{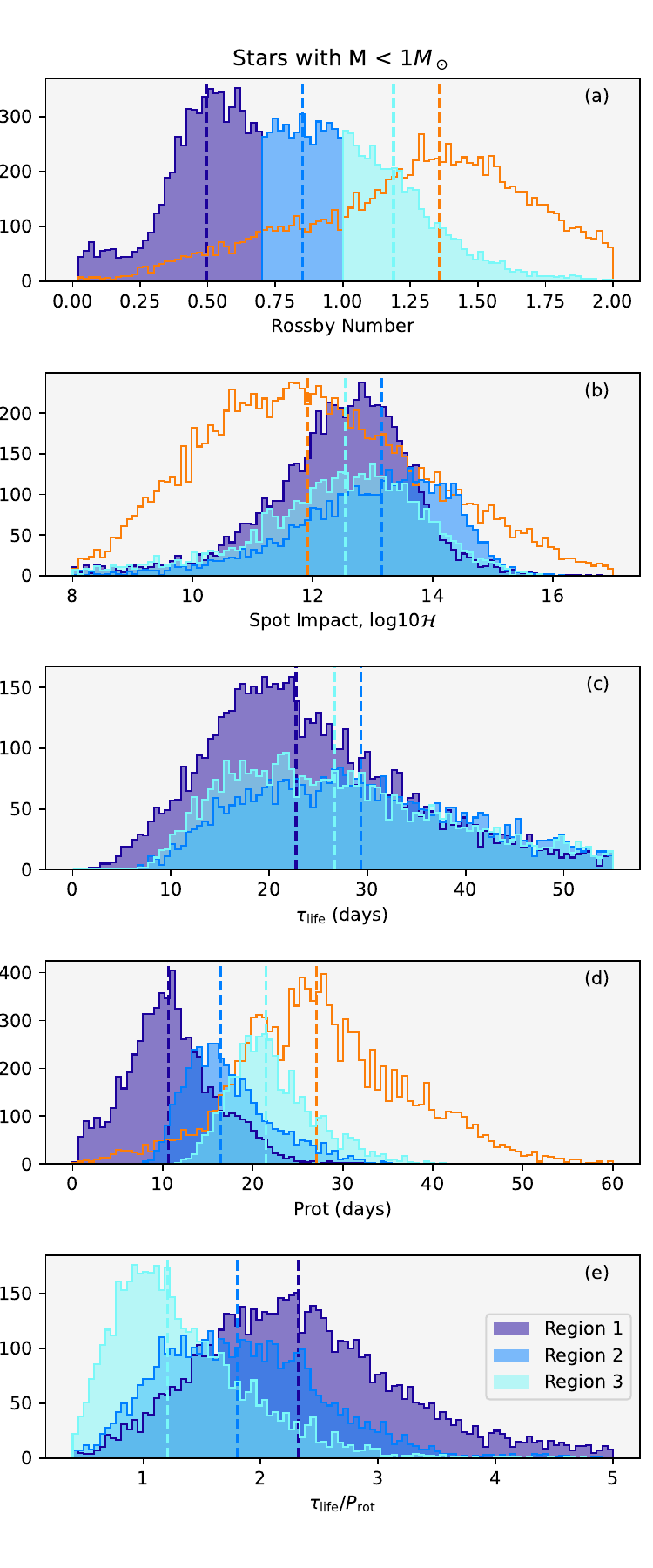}
      \caption{In blue: the sample of stars with rotation peaks. The different shades of blue correspond to the three regions of Fig. \ref{fig:lnh_mck} cut   according to the Rossby number. This slicing is illustrated in panel (a) as a function of the Rossby number. 
      The orange histograms represent the peakless group. 
      From top to bottom, the histograms are for (a) the Rossby Number, (b) spot impact proxy, (c) the mean lifetime $\life$,(d) the rotation period and finally (e) the ratio $\prot / \life$.}
      
         \label{fig:hist_type}
\end{figure}

As seen in Sect.\,\ref{subsec:lifetime_mck} and \ref{subsec:spot_impact_result} the main proxies ($\lnH$ and $\life$) used to characterise spot signature in LCs depend on stellar properties such as mass, $\prot$ and Rossby number, which we use to define stellar categories.

First, as seen in Fig.\,\ref{fig:lnh_mck}, stars with M>1$\rm M_\odot$ show \correct{smaller} values of $\lnH$ regardless of other stellar properties, whereas stars with M<1$\rm M_\odot$ show \correct{larger} values of $\lnH$ and $\life$ presenting a clearly structured behaviour.
We thus focus on stars of lower mass, which are in addition more numerous than the more massive stars.

The histograms of Fig. \ref{fig:hist_type} summarize the different trends of this population, providing a clearer illustration of their statistical distributions.
The Rossby number being a fundamental index when dealing with magnetic activity, we use it to define categories of stars depending on its value and Fig.\,\ref{fig:lnh_mck}: Region 1 with $\rm Ro<0.7$, Region 3 with $\rm Ro>1$, and Region 2 between these two values. The distributions of stars for these regions is shown in panel a of Fig.\,\ref{fig:hist_type}, as well as the distribution of peakless stars (in orange) having clearly higher $\rm Ro$. This finally leads to four categories of stars:

\begin{itemize}
\item Panel b: spot impact $\lnH$.

These four categories show wide distributions of spot impact $\lnH$ but with different median values, Region 2 having the highest median while Regions 1 and 3 have lower (and almost equal) medians, and peakless stars have the lowest median spot impact.

\item Panel c: lifetime proxy $\life$.

Lifetimes ($\life$) of the stars from the 3 regions (lifetimes are not defined for peakless stars) also show wide distributions but again with different median values: Region 1 being the shortest, Regions 2 and 3 being longer (panel c of Fig.\,\ref{fig:hist_type})
\correct{
Regarding the bias towards long lifetimes for slow rotators, the distribution in Region 1 is likely the closest to the underlying distribution, while the other two regions are increasingly biased. However, examining the median values reveals that stars in Region 2 tend to have slightly longer lifetimes than region 3. This suggests that Region 2 may be less affected by methodological bias, with its distribution being closer to the underlying one.
Additionally, panel (e) shows that stars in Region 2 have a $\life / \prot$ ratio close to 2, whereas stars in Region 3 exhibit a ratio closer to 1. This latter group, which has shorter lifetimes, is much more likely to be biased towards long lifetimes. }

As mentioned before, we do not have lifetime estimations for the peakless group of our sample but the rotation period gives the upper limit of their characteristic lifetime. 
Fig. \ref{fig:hist_type}d shows that these stars have much slower rotations than the estimate mean lifetime of Region 2 and 3; they should therefore have shorter or equivalent lifetimes.

\item Panel d: rotation period.

Panel \ref{fig:hist_type}d shows the rotation period of the three different regions. 
As they are based on the Rossby number, we can see that each of them corresponds to different rotation intervals that increase with their Rossby value. 
We observe that stars with the fastest rotation exhibit the shortest lifetimes (see Fig. \ref{fig:life_prot}).

\item Panel e: ratio $\life/\prot$.

Because the $\prot / \life$ ratio is the parameter governing the appearance of rotation peaks, it is also interesting to represent its distribution in Fig. \ref{fig:hist_type}e. 
These distributions of $\life$/$\rm \prot$ show quite different distributions than $\life$ alone: Region 3 with the longest $\prot$ has the smallest $\life$/$\rm \prot$ and Region 1 with the shortest $\prot$ has the largest $\life$/$\rm \prot$.
In fact, this means that even if fast rotators have the shortest lifetimes , the spots of these stars transit over the stellar disc several times, creating clearly visible rotation peaks. It is precisely for this reason that fast-rotating stars often have better defined rotation peaks than slower-rotating stars like the Sun. Similarly, less visible rotation peaks in slower-rotating stars are not necessarily associated with shorter spot lifetimes.
\end{itemize}

Based on Fig.\,\ref{fig:lnh_mck} and \ref{fig:hist_type}, the observed stars and their type of activity can be described as follows:

\begin{itemize}
    \item Activity of stars under $1\rm M_\odot$ show trends depending on the Rossby number.
    When considering the median value $\rm Ro$ of the distribution shown in
    Fig.\,\ref{fig:hist_type}, these trends can be described as:
    \begin{itemize}
        \item $\rm Ro < 0.7$ (Region 1): these stars have intermediate values of spot impact and relatively short lifetimes, however increasing with decreasing mass.
        These stars correspond to the fast rotators branch of \cite{McQ2014}.
        \item  0.7\,<$\rm Ro < $\,1 (Region2): it corresponds to the most active stars with the highest spot impact and also the longest lifetimes.
        \item $\rm Ro > 1$ and $\rm Ro < 1.5$ (Region 3): these stars have intermediate spot impacts as Region 1 and lifetimes shorter than those of Region 2 but longer than those of Region 1. 
        
        \item $\rm Ro > 1.2$: it is the case of a large fraction of stars showing peakless
        spectra: they have also smaller spot impacts than the 3 other groups depicted before, which suggests a fourth type of activity.
        If we look at their rotation period, these stars are likely to have shorter mean lifetimes than the groups of stars discussed above. 
        
        \end{itemize}
    \item For more massive stars ($>1 \rm M_\odot$), the behaviour differs. Spot impact and lifetimes are significantly lower compared to lower mass stars, indicating that they are considerably less active. Additionally, stars with peakless spectra display similar low activity levels.
    This reduced activity in massive stars is not unexpected: these stars correspond to masses where the convective zone becomes thinner, resulting in a less efficient Parker-type dynamo effect.
\end{itemize}

\section{Discussion}
\label{sec:disussion}

The main results of this work are now discussed with regard to previous studies on stellar activity and recent observations of magnetic fields by Zeeman-Doppler imaging. We address the potential insights brought by this work to stellar internal magnetism, and finally discuss the physical phenomena that can bias our methodology.
\begin{figure*}
   \resizebox{\hsize}{!}
            {\includegraphics[]{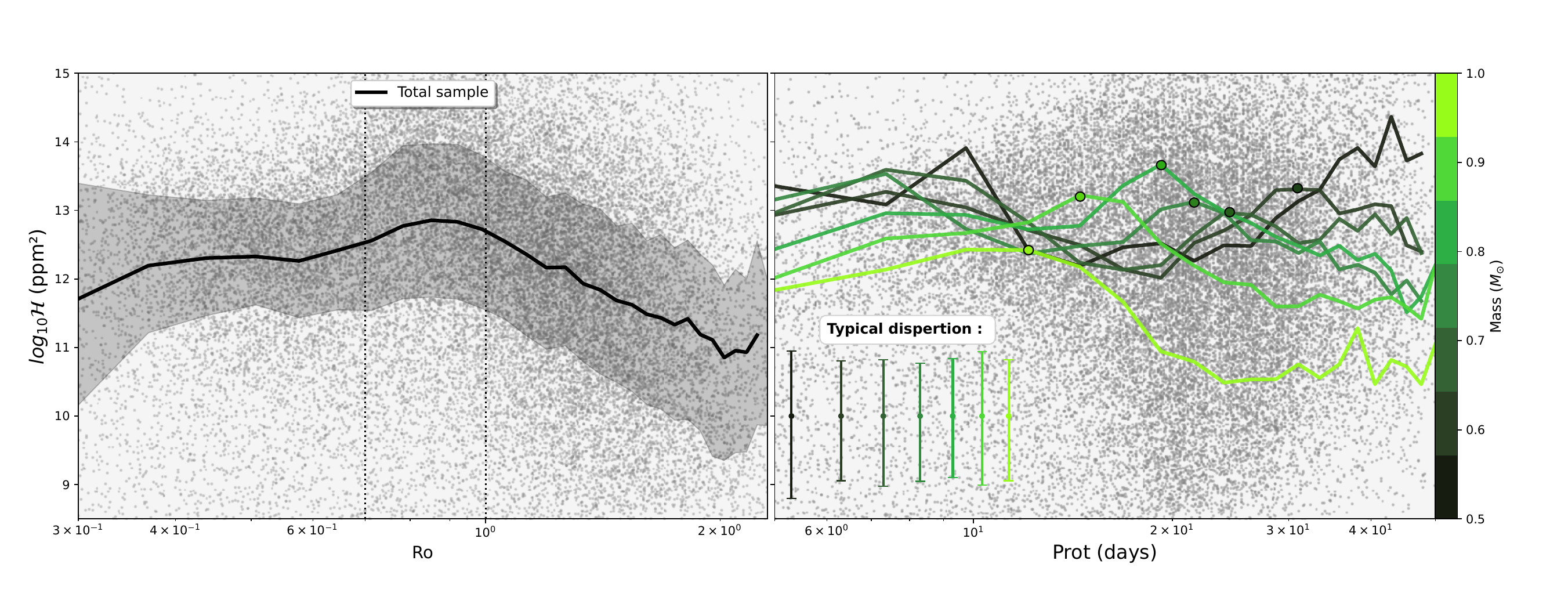}}
     \caption{Scatter plot of the total sample of stars (grey dots in the background) as a function of the spot impact proxy and the rotation Rossby number (left) or period (right).   
     On the left panel, the black line represents the median value of $\lnH$ as a function of $\rm Ro$ for the whole sample. The hatched area around this line represents the interval of the quartiles.
     The dotted lines show the position of $\rm Ro = 0.7$ and $1$.
     For the right panel,
     these stars are grouped into slices of increasing masses (from $0.5 \, \rm M_\odot$ in dark to $1 \, \rm M_\odot$ in light green). The coloured lines represent the median values of $\lnH$ for each of these mass samples. 
     The typical on $\lnH$ have been represented at the bottom left of the panel. They have been estimated by computing first quartiles around the median value. 
     The large coloured dots mark the position of the maxima of these curves. 
     }
        \label{fig:slices_mck}
  \end{figure*} 
\subsection{Comparison with other activity proxies}

Two other common activity proxies have been studied in the past, namely the X-ray luminosity and the mean level of photometric variability $S_{\rm ph}$.

\cite{Wright2011} studied the relation between stellar activity in X-rays and rotation period in a sample of about 800 low-mass stars. Fig. 2 of \cite{Wright2011} in particular shows a scatter plot from this sample of stars as a function of the Rossby number (from $\rm Ro = 10^{-3}$ to $ 3$) and the X-ray luminosity. This scatter plot shows a very clear plateau for $\rm Ro$ values lower than $\sim 0.1$ and a decrease above.

We have reproduced the same figure, plotting $\lnH$ as a function of the stars' rotation Rossby number (left) and period (right) in Fig. \ref{fig:slices_mck}.


Our sample covers a slightly narrower range of Rossby numbers between 0.5 and 4, which gives us only a small overlap with the data from \cite{Wright2011}. However, we clearly do not retrieve similar trends in that range of values. Indeed, the impact proxy $\mathcal{H}$ exhibits instead a maximum between $\rm Ro = 0.7$ and 1 (see the black line of Fig. \ref{fig:slices_mck}). The difference between these two trends could be due to the difference in the active regions probed by each indicator. In the present work, we focus on the signature of spots, which are mostly correlated with the photosphere properties, while X-ray luminosity is more sensitive to coronal features. Moreover, it is also worth mentioning that we use a larger sample of stars than \citet{Wright2011}, possibly allowing a better mapping of the trends with the stellar parameters.

For the right panel of Fig. \ref{fig:slices_mck} we choose to represent the median value  of $\lnH$ for increasing intervals of masses from 0.5 to 1 solar masses.  As illustrated in Fig. \ref{fig:lnh_mck}, $\lnH$ exhibits distinct behaviours depending on both mass and rotation. The higher masses are not represented, as they show no particular structure.
When considering the slowest rotators, a series of maxima appears (large dots). These maxima amplitudes increase with mass until it reaches its largest value for the mass interval $M = 0.8\,\rm M_{\odot}$. 
The value of the maxima then goes down.
These maxima also depend on the rotation period: for low masses intervals, the maximum corresponds to a rather slow rotation, while for more massive stars in this subsample it corresponds to smaller rotation periods.
For stars with very low mass ($ < 0.8 \rm M_{\odot}$), a second regime of high $\lnH$ seems to appear for the faster rotating stars. 
However, this second activity regime remains more uncertain due to the reduced amount of data that we have for low masses in this region.

On the other hand, the left panel of Fig. \ref{fig:slices_mck} can also be compared to the photometric activity index $S_{\rm ph}$ that was measured in a similar stellar sample by \cite{2024Santos}; the result is plotted in Fig. 1 of this latter paper as a function of the rotation period and for different spectral types, which are also representative of different mass ranges. 
The results of \cite{2024Santos} show a shape similar to our results, when considering the global scatter plot. 
This is not surprising, as the impact proxy $\mathcal{H}$ is defined as the mean level of variability due to spots in the considered light curve, and it is thus  similar to  $S_{\rm ph}$ index. However, if you look at the trends by mass (or by spectral type in \cite{2024Santos}), you do not see a maximum of activity as a function of rotation, just a plateau. 
This can be attributed to our choice of representation of our results, but also because our 
model in the Fourier domain allows us to retrieve additional information, namely the mean spot lifetime. This new index helped us to further characterize the stellar activity compared to these previous studies by identifying different activity regimes as a function of the Rossby number, clearly demonstrating the value of our new modelling.

\subsection{Link with the magnetic field topology}

Thanks to recent spectropolarimetric observations (e.g. ESPaDOn, NARVAL), the three-dimensional configuration of surface magnetic fields could be reconstructed for a few stars using Zeeman-Doppler techniques \citep[e.g.][]{Donati2009,2023Donati}. As seen in Fig. 3 of \citet{Donati2009}, different trends emerge, based on the iso-Rossby lines : stars with $\rm Ro > 1$ have a weak magnetic field with a poloidal axisymmetric topology while stars with $0.1 \lesssim \rm Ro \lesssim  1$ have stronger, toroidal and non axisymmetric magnetic field. This sample is composed of about 15 stars with masses comparable to those in our stellar sample, that is, between about $0.5$ and $1.2$~M$_\odot$; stars with $\rm Ro \lesssim 0.1$ have generally very strong poloidal fields, but correspond to stars with mass under $0.5~\rm M_{\odot}$, which are outside the range of our study sample.

To put these observations in perspective with our study, we choose in Fig. \ref{fig:lnh_mck} similar axes to Fig. 3 in \cite{Donati2009}.
Our results only cover the stars from 0.4 to 1.4 solar masses, which correspond to the top of the Fig. from \cite{Donati2009}
According to this figure, and focusing on masses below about 1.2\,$\rm M_{\odot}$ as in \cite{Donati2009},  
we see that stars with $\rm Ro < 0.7$ (Regions 1) have short $\life$ and intermediate spot impact $\lnH$ for weak Rossby number, then reach a regime of intense activity with a large spot impact and long lifetimes (region 2). 
For stars with $\rm Ro > 1$ (Region 3 and peakless group), we again observe a medium and even low impact spot for the highest Rossby numbers.  Lifetimes are longer than those below $\rm Ro = 0.7$.

We can therefore establish links between these proxies and the topologies described by \cite{Donati2009}. Firstly, short lifetimes seem to appear for toroidal and asymmetrical topologies, while long ones appear for more stable and poloidal configurations. Then,
it is interesting to note that the transition at $\rm Ro\approx 1$ between the two kinds of observed topologies is near the high activity regime. Indeed, $\rm Ro \approx 1$ indicates a change of physical regime and the high observed activity index could be the signature of a high variability in the stellar magnetic structure (e.g., intermittent reversals) at the transition between two stability domains. 

The relation depicted between the activity properties revealed in this work for a large sample of stars and the observed magnetic fields for a handle of close stars remains preliminary. In particular, the question of the variability over time and of the magnetic cycles is certainly of prime importance to perform relevant comparisons. 
Regarding magnetic cycles, \citep{See2016, Fabbian2017} have produced the same plot as \cite{Donati2009}, but have connected the magnetic field topologies to the magnetic cycles by looking at the active and inactive branches of \citet{Saar1992, Branderburg1998}. 
They have also showed the existence of a delimitation between the two branches at $\rm Ro = 1$ between the long and short cycles, as it was observed in \cite{Saar1992}.

\subsection{Insights into the evolution of stellar dynamo}

As seen in the previous section, different observations show that the magnetic properties of stars depend on the Rossby number. In particular, a transition seems to operate in the spot activity, field topology \citep{Donati2009} or magnetic cycles \citep{Saar1992, See2016} around $\rm Ro = 1$. In this work, we have shown that this transition is predominantly a maximum in the impact proxy, as well as the progressive emergence of long-lived spots as Ro increases and the decrease of the $\life/\prot$ ratio.
These changes of regimes as a function of the Rossby number are also supported by 3D MHD numerical simulations that have already highlighted change of regimes with the variation of the Rossby number in the differential rotation \citep[e.g.,][]{Brun2017}, dynamo processes \citep[e.g.,][]{Raynaud2015,Zaire2022} and 
cycles \citep[e.g.,][]{2022Brun}. 
 \begin{figure}
   \centering
   \includegraphics[width=\hsize]{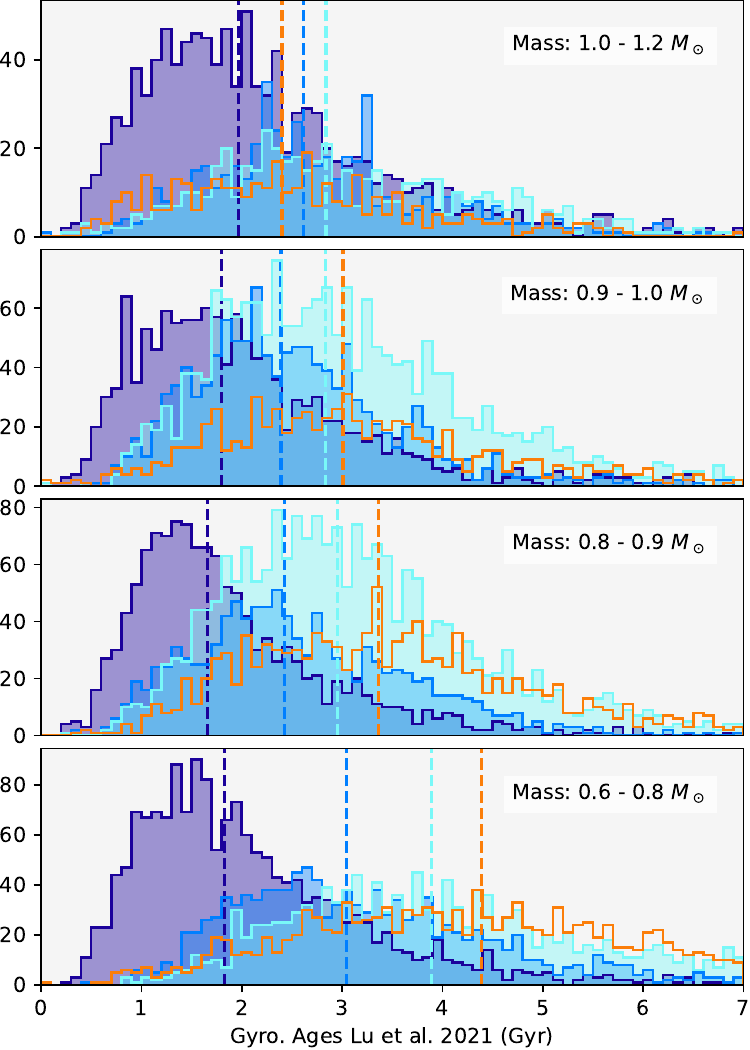}
      \caption{Distribution of star ages for increasing mass intervals. Each row corresponds to a decreasing mass interval (from $1$ to $1.2M_\odot$ for the first row, $0.9$ to $1.0M_\odot$ for the second row, $0.8$ to $0.9M_\odot$ for the third row and $0.6$ to $0.8M_\odot$ for the last row). The colours of the histograms correspond to the different regions of Figure \ref{fig:lnh_mck}. For each of these histograms, a line of the same colour is drawn to represent the median of the histogram.}
      
         \label{fig:age_hist}
   \end{figure}

Although good guides, current simulations are nevertheless still far from realistic regimes, and the constraints such as those brought in this work represent a good opportunity to test their theoretical outcomes. For instance, one could be tempted to draw a simple picture for the evolution of the dynamo mechanisms according to the simple trends observed as a function of the Rossby number. Indeed, owing to the efficient magnetic wind breaking in such low-mass stars, the Rossby number is expected to increase as they evolve on the main-sequence \citep[e.g.][]{2024Noraz}.  
If we take into consideration the possible link between the Rossby number and the age of the stars.

A catalogue of \textit{Kepler} stars with ages has been published by \cite{2021Lu}, based on a gyro-chronological study. Fig. \ref{fig:age_hist} shows the evolution of these ages for each intervals of Rossby defined on Fig. \ref{fig:lnh_mck}. 
This figure was made for different mass ranges, since the time they spend on the main-sequence are not the same depending on their mass. 
For stars with masses below one solar mass, the median values shows that the Rossby intervals do indeed increase with age. This suggests that a star may transition through the various types of activity outlined in Sect.\ref{subsec:summary_result}. However, it is important to note that the distributions are quite broad and show significant overlap, particularly among stars with peakless spectra and those in Regions 2 and 3. It could be explained by an incorrect estimation of the ages, especially given that age estimation through gyrochronology is not a completely robust method yet, especially for high masses, or a bad sorting of the spectra with or without peaks.
For stars with masses $> 1\rm M_\odot$, the position of the median value of the peakless group became lower than Regions 2 and 3 and the different regions are even more spread out.

In this way, Fig.\,\ref{fig:slices_mck} could suggest that the activity type evolves with time at a given mass. For low masses, stars are likely to evolve in a constant phase with moderate activity, followed by an increase of activity near $\rm Ro = 1$ and then a rather rapid decrease. For stars with masses above $1\rm M_{\odot}$, the stars potentially go from the activity of Region 1 (moderate $\lnH$ and short $\life$) to low $\lnH$ and longer $\life$, without the “bump” phase.

\subsection{Limitations of the model: Influence of faculae and differential rotation}
\label{sec:facula_dr}

The method used in this paper is based on an analytical model detailed in Sect.\ref{sec2} where two physical phenomena have been neglected: the differential rotation and the appearance of faculae on the stellar surface. 
The goal of this section is to discuss the validity of these assumptions regarding the results presented in this paper. 

\subsubsection{Influence of faculae}

Faculae are bright magnetic regions that appear on the stellar surface, thus being able to contribute to photometry by inducing an increase in luminosity, unlike spots. They have been observed and characterised well on the Sun, but their impact and presence on other stars are still current topics of research. 

Their effective contribution is debated. \cite{Basri2018} explains that their impact might not be readily discernible in \textit{Kepler} data if their influence is supposed to be similar to that of the solar faculae.
However, the analysis by \citet{Shapiro2017} shows that the amplitude of the rotation peaks is diminished by faculae in the solar case, which may influence the estimate of $\lnH$ but not the width of the rotation peaks. 
However, if we go back to the description of our model, the impact of faculae should mainly be seen in the transit-dependent part of the spectrum. Indeed, \cite{2020Toriumi} explains that the impact of faculae on brightness results in an increase in photometry at the edges of the transit function. In this case, it is the characteristic transit time and possibly $\lnH$ that are affected. 
The reason why the rotation peaks of the Sun are very barely visible is because it has mainly spots with lifetime close to its rotation period.

Besides, \cite{Reinhold2019} studied the distinction between spot- and faculae-dominated stars using photometric and chromospheric data. This work showed that the faculae-dominated stars have Rossby number above 1 and $\prot > 20$ days.
This corresponds to the area where the stars with peakless spectra are dominant.
This might seem contradictory to our previous argument. However, these stars exhibit the slowest rotation rates, making their rotational peaks more challenging to detect. The visibility of such peaks depends on the number of transits a spot can produce during its lifetime, and slower rotators naturally have fewer observable transits.

\subsubsection{Influence of the differential rotation}

The DR corresponds to a latitudinal shear between the pole and the equator. Since all the spots are not transiting at the same latitude, and therefore may have a different rotation period \citep{1859Carrington,Reinhold2013}, the DR can widen rotation peaks and thus induce a bias on the spot lifetime proxy.

\begin{figure}
   \centering

   \includegraphics[width=\hsize]{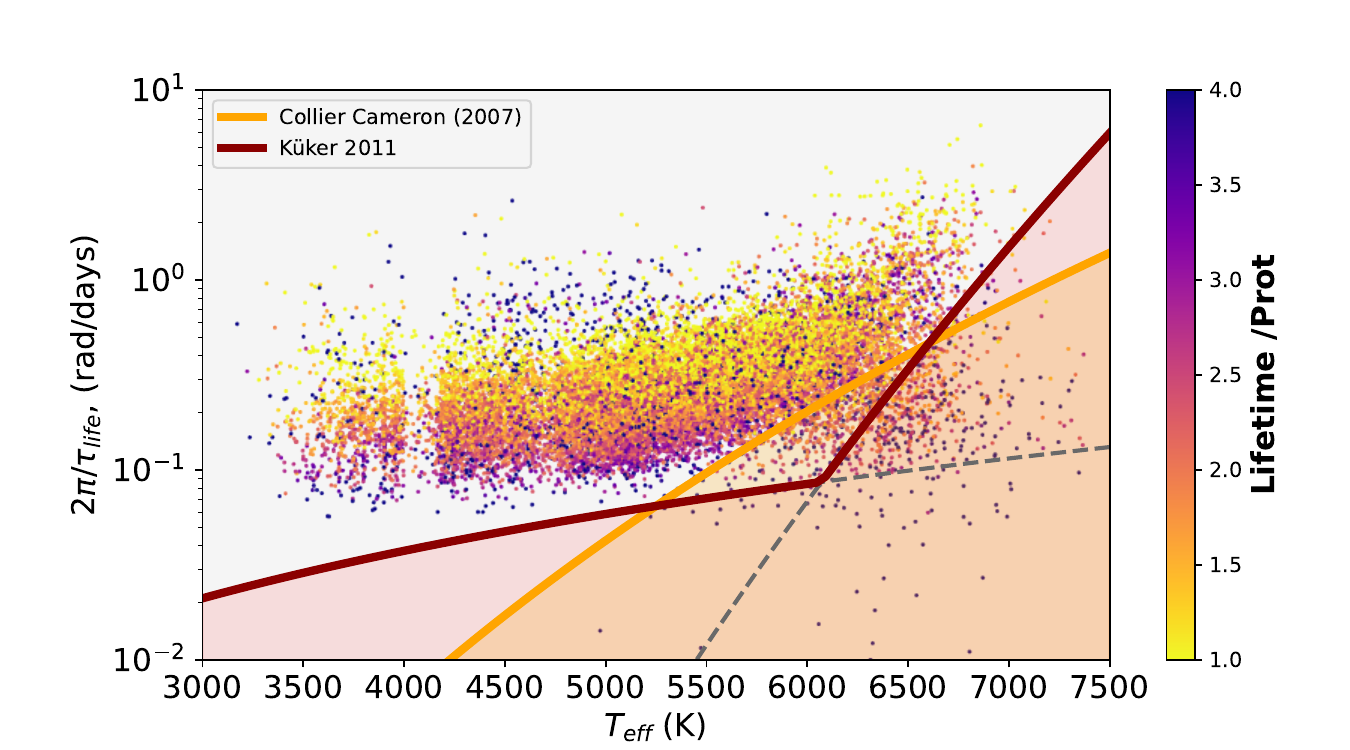}
      \caption{Spot lifetime estimation expressed in radians per days as a function of the effective temperature in K. The orange line is the $T_{eff}-\Delta \Omega $  relation proposed by \cite{Collier2007}, with $\Delta \Omega$ in days. The dark red line is the relation from \cite{Kuger2011}, with the dashed lines representing the two components of this relation. 
      The hatched regions are the ones where Eq. \ref{eq:dr_condition} is not verify, i.e. the DR may affect the lifetime estimation, $\life$. 
              }
         \label{fig:DR}
   \end{figure}
The difference with the model presented in Sect.\ref{subsec:n_spots}
is that each spot is now associated with a Dirac comb, with a spacing between the values of the rotation period at the equator and at the pole.
In the Fourier domain, we therefore have a superposition of comb-like signals shifted in frequency from each other by $n \Delta \Omega$, with $\Delta \Omega$ the value of the DR relatively to the equator at the latitude at which the spot has transited and $n$ the number of the harmonics considered.
This shift in frequency contributes to broadening the rotational peaks.
Thus, the width of the $n$-th rotation peak is dominated by the spot lifetime if: 
\begin{eqnarray}
    n\Delta \Omega < \frac{2\pi}{\life}
    \label{eq:dr_condition}
\end{eqnarray}
Thus, the higher harmonics can be widened by the DR. Since no more than three rotation peaks have been observed in the \textit{Kepler} data and that the algorithm fits only two of them, this potential bias is not distinguishable from noise. 
 
Are there some stars where the DR can be observed in the first harmonics? 
Multiple DR estimation have been provided in the past years using different methods. 
Applying Doppler imaging, \cite{Barnes2005} found a DR that increases with the effective temperature. 
Combining these results with DR estimates through the broadening of the rotation peaks in the Fourier spectrum by \cite{Reiners2006}, \cite{Collier2007} proposed a power-law fit of the form $\Delta\Omega = 0.053( \teff /5130)^{8.6}$. 
Later, based on stellar models, \cite{Kuger2011} showed that $\Delta \Omega$ depends on two power laws, depending on the range of temperature of the stars. 
The broadening of rotation peaks in the power spectra have also been used to determine the DR in the \textit{Kepler} data by \cite{Reinhold2013, Reinhold2015}. 
However, this method has been questioned by \cite{Basri2020}, whose results show that DR is only visible for very long lifetimes ($ \sim 10 - 15 \prot$). 
In this work, the temporal evolution of the spots is modelled by a symmetric and triangular function and their lifetime is defined as the total width of this function. Taking the width at half height as in our formulation, their result translates into a ratio $ \life / \prot $ between 4 and 6 in our model.

Fig. \ref{fig:DR} represents the estimation of the spot lifetime of this paper as a function of the effective temperature. We add the $\Delta\Omega-\teff$ relations from \cite{Collier2007} and \cite{Kuger2011}. 
In these studies, the stars that do not respect the criteria of Eq.\,\ref{eq:dr_condition} are those with long lifetimes and mostly high temperatures and represent a very small fraction of the sample. 

\section{Conclusions}
\label{sec:conclusion}

Stellar spots and their photometric signature represent very interesting probes of stellar magnetism for large samples of stars. However,
finding information about spot properties in LCs is a complex process due to the intrinsic degeneracies of the problem. Indeed, the impact of a spot on a light curve depends on many of parameters such as its position, size, temperature, the rotation axis inclination of the star, and even its differential rotation.
As a result, different stellar configurations and spots distributions and properties can lead to similar impacts on LC in the temporal domain.

In this paper, we proposed a new analytical model of the LCs PSD in the Fourier domain that allows us to lift some of these degeneracies while accounting for the main properties of spots.
This model takes into account the rotation peaks of the LCs power spectrum and other components of the spectra at low frequencies. 
This LC power spectrum model introduces three proxies to estimate average spot properties: $\ttransit$, $\life$, and $\lnH$.
This description shows that rotation peaks in the PSD occur when stars have spots with lifetimes exceeding the rotation period. Consequently, two spectral categories emerge: those dominated by spots with multiple transits that have apparent rotation peaks harmonics, and those from stars with a majority of spots that create a single transit in the LC, which thus have a peakless spectrum (like the Sun). 
After validation by simulations, we applied this model to a large \textit{Kepler} dataset consisting of the catalogues of \cite{McQ2014} and \cite{Santos2019, Santos2021}.  

The data analysis yields several results. 
We show a clear distinction in the spot activity according to the stellar mass in the \textit{Kepler} data. 

Stars above one solar mass have a much lower spot impact than low-mass stars. For stars between $1$ and $1.2 \rm M_\odot$, lifetimes also increase with Rossby number.
For stars below $1 \rm M_\odot$, the proxies adopt different values depending on the Rossby number. In particular, an intense activity regime is observed for stars between $\rm Ro= 0.7$ and $1$. Spot lifetimes also tend to increase with the star's rotation period. Increase that seems to stop or even decrease a little after $\rm Ro =1$. Since the peakless spectra are mainly present for large Rossby values, the evolution of spot lifetimes is difficult to infer for these values.
For $\lnH$, a decrease is noted beyond $\rm Ro = 1$. 
These variations indicate that the evolution of spot activity with Rossby number is not as linear as generally suggested in the literature.

These various types of activity described in terms of Rossby number indicate a potential connection to distinct dynamo regimes.
They can further be correlated with the topology map constructed by \cite{Donati2009}.
Moreover, stars in the high activity regime have Rossby numbers close to 1. Previous studies have indicated behavioural changes near this Rossby number, such as alterations in magnetic cycles and topology. This makes the observed high spot activity near this delineation particularly interesting.

This new methodology itself has potential for further development, with the possibility to study the impact of DR or use transit time estimates, which may be related to the star's inclination or the appearance of faculae.

We conclude that these results offer promising insights that could shed light on understanding the inner dynamo of stars.
We plan to correlate these results with simulations of stellar interiors to gain a deeper understanding of the three distinct regimes we observed. 
A comparison with the work of \cite{2022Basri} on the spot lifetimes is also planned. 
We also plan to put into perspective our results with the LAMOST activity observations. 
In the longer term, these results could aid in the interpretation of PLATO data, particularly for the fitting algorithm that might be used on future light curve power spectra. They would also help to better constrain the impact of stellar spots on photometry and the information they provide about stellar magnetism.

\begin{acknowledgements}
     This paper includes data collected by the Kepler mission and obtained from the MAST data archive at the Space Telescope Science Institute(STScI). Funding for the Kepler mission is provided by the NASA Science Mission Directorate. STScI is operated by the Association of Universities for Research in Astronomy, Inc., under NASA contract NAS 5–26555.
     This work has made use of data from the European Space Agency (ESA) mission
{\it Gaia} (\url{https://www.cosmos.esa.int/gaia}), processed by the {\it Gaia}
Data Processing and Analysis Consortium (DPAC,
\url{https://www.cosmos.esa.int/web/gaia/dpac/consortium}). 
     We warmly thank Cilia Damiani to provide us the simulation code to create light curve. 
     We also thank K. Belkacem, A. S. Brun, N.Meunier, P. Boumier and L. M. R. Lock for useful discussions.
     We also warmly thank the anonymous referee for all his comments and thoughtfulness, as they have enabled us to significantly improve our article.
          
\end{acknowledgements}

\bibliographystyle{aa}
\bibliography{biblio.bib}

\begin{appendix} 
\begin{figure*}

\resizebox{\hsize}{!}
            {\includegraphics[width=\hsize]{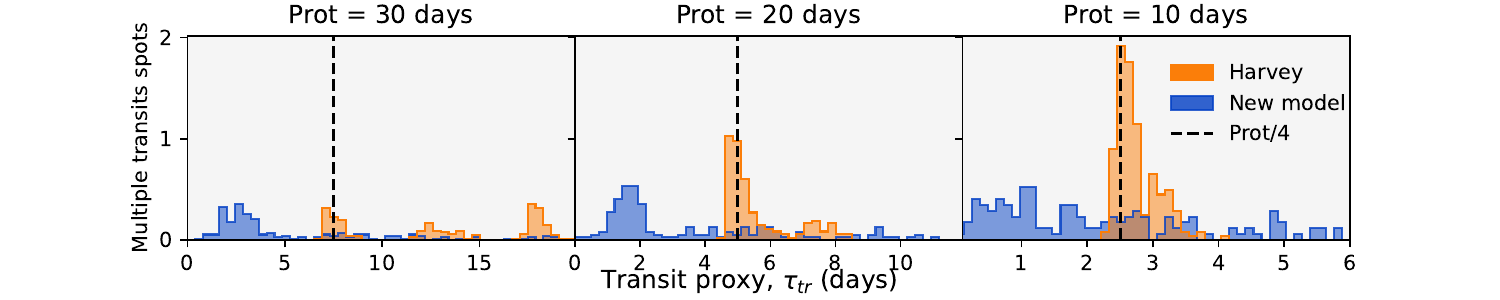}}
     \caption{ Same as in Fig. \ref{fig:tau_hist}, but for the simulations with only multiple transits spots.
             }
        \label{fig:hist_tau_multi}
  \end{figure*}
\section{Transit time in the case of the multiple-transit spots simulations}\label{appendix:longspot_simu}

Fig. A\ref{fig:hist_tau_multi} shows the results obtained for the mean transit time on light curves simulated with multiple-transit spots only. 
The fitted value is underestimated and much more widely dispersed for our model than for the Harvey profile, which was expected.
In this case, there are only spots with multiple transits and therefore no contribution from those with only a single transit, making it impossible to observe this proxy.
More precisely, the new model tries to fit a $\rm S_{single}$ contribution while it is not present in the simulated data for only multiple-transit spots.
This component converges towards values close to $\prot$/8, which correspond to the third rotation peaks counting from the fundamental, since only the first two are fitted by the rotation peaks component. In fact, Fig. \ref{fig:PSD_model} shows that the single-transit spot component converges on this third peak.

\begin{figure}
  \centering
  \includegraphics[width=\hsize]{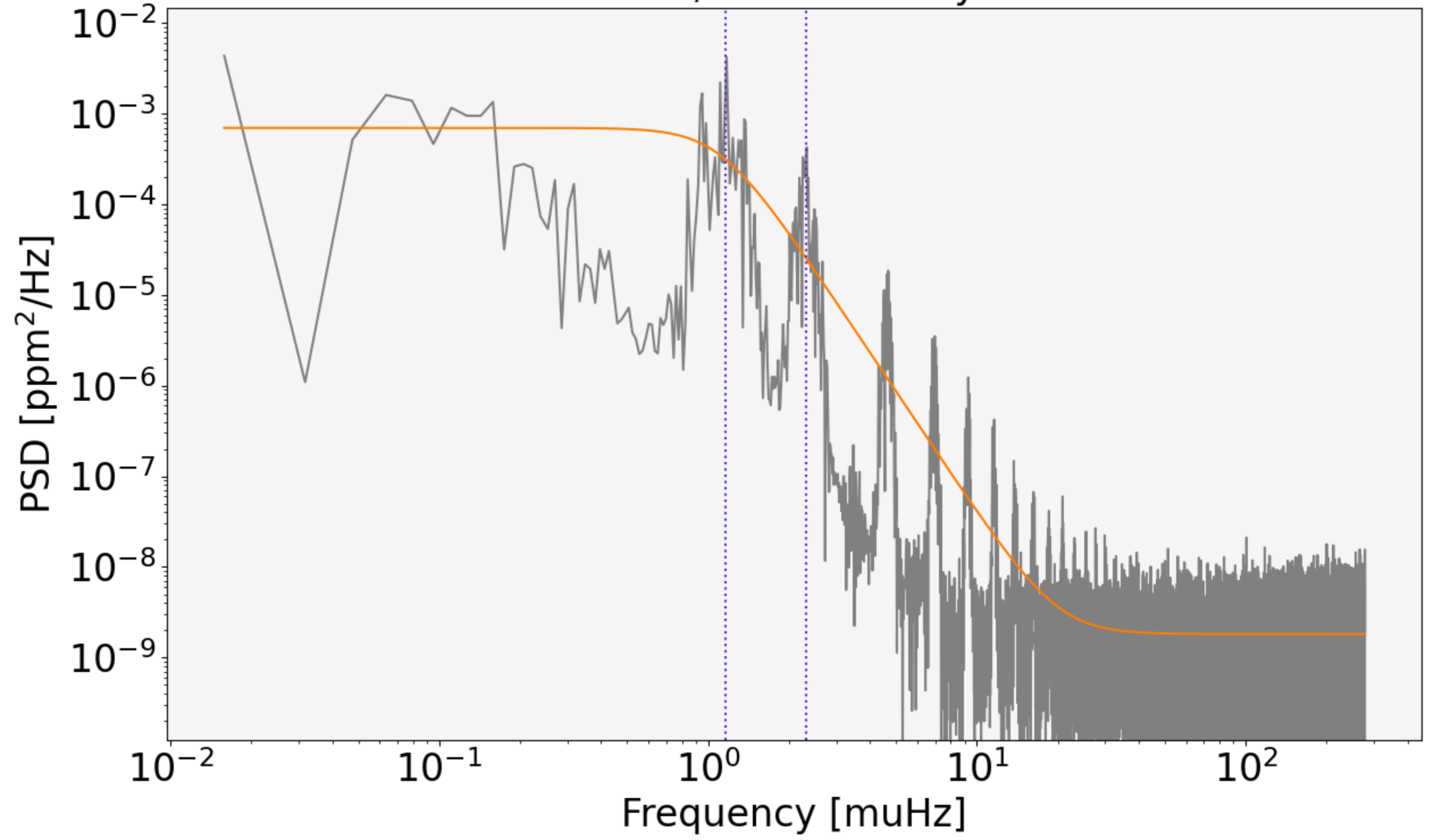}
     \caption{ Power spectrum of a simulate LC with only multiple transit spots. The spectrum has been fitted with the Harvey model (blue).
             }
        \label{fig:PSD_harvey}
\end{figure}

\begin{figure}
  \centering
  \includegraphics[width=\hsize]{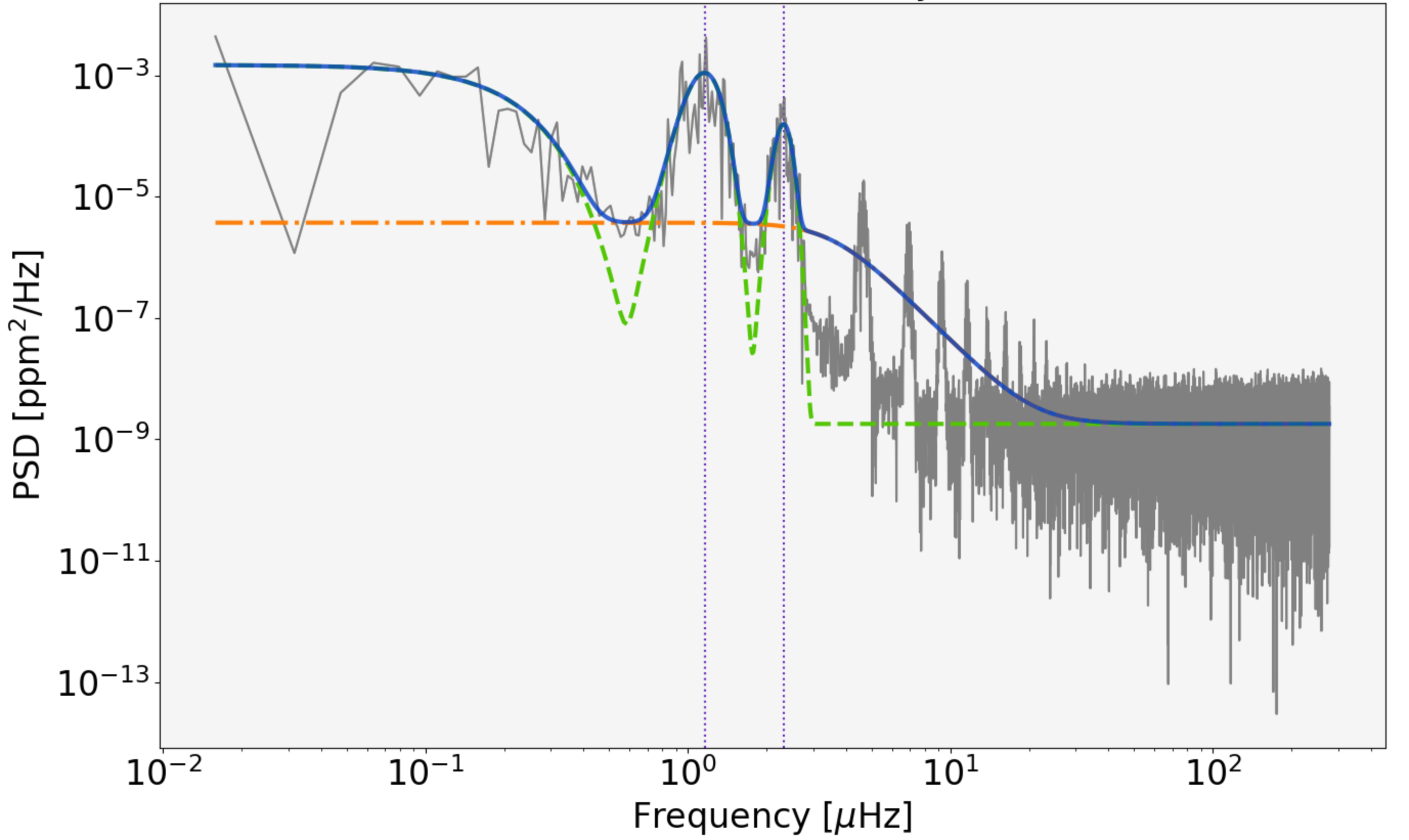}
     \caption{Same Fig. as Fig. \ref{fig:PSD_harvey} but fitted with the new model describe in this article. 
             }
        \label{fig:PSD_model}
\end{figure}

For the Harvey profile, the median value of the distribution is close to the expected one.
It is also an expected result: because there is many rotation peaks, the fit is  driven by their amplitude (see Fig. \ref{fig:PSD_harvey}), which is linked to the transit profile (see Fig. \ref{fig:model_ana}).
There are also several of accumulation points on higher values, for the same reason that for the mixed-spots cases: in some cases, when only one or two rotational harmonics are observable, the fit with the Harvey profile underestimates the  width at half maximum.

\section{Technical description of data cleaning}
\label{appendix:technical_algo}

As a first step and to simplify the methodology, the Harvey profile and our model were applied to all the data. The classification of spectra into those with or without peaks was performed subsequently, as the algorithm for this distinction was developed at a later stage of the study.

\begin{figure}
\resizebox{\hsize}{!}
            {\includegraphics[width=\hsize]{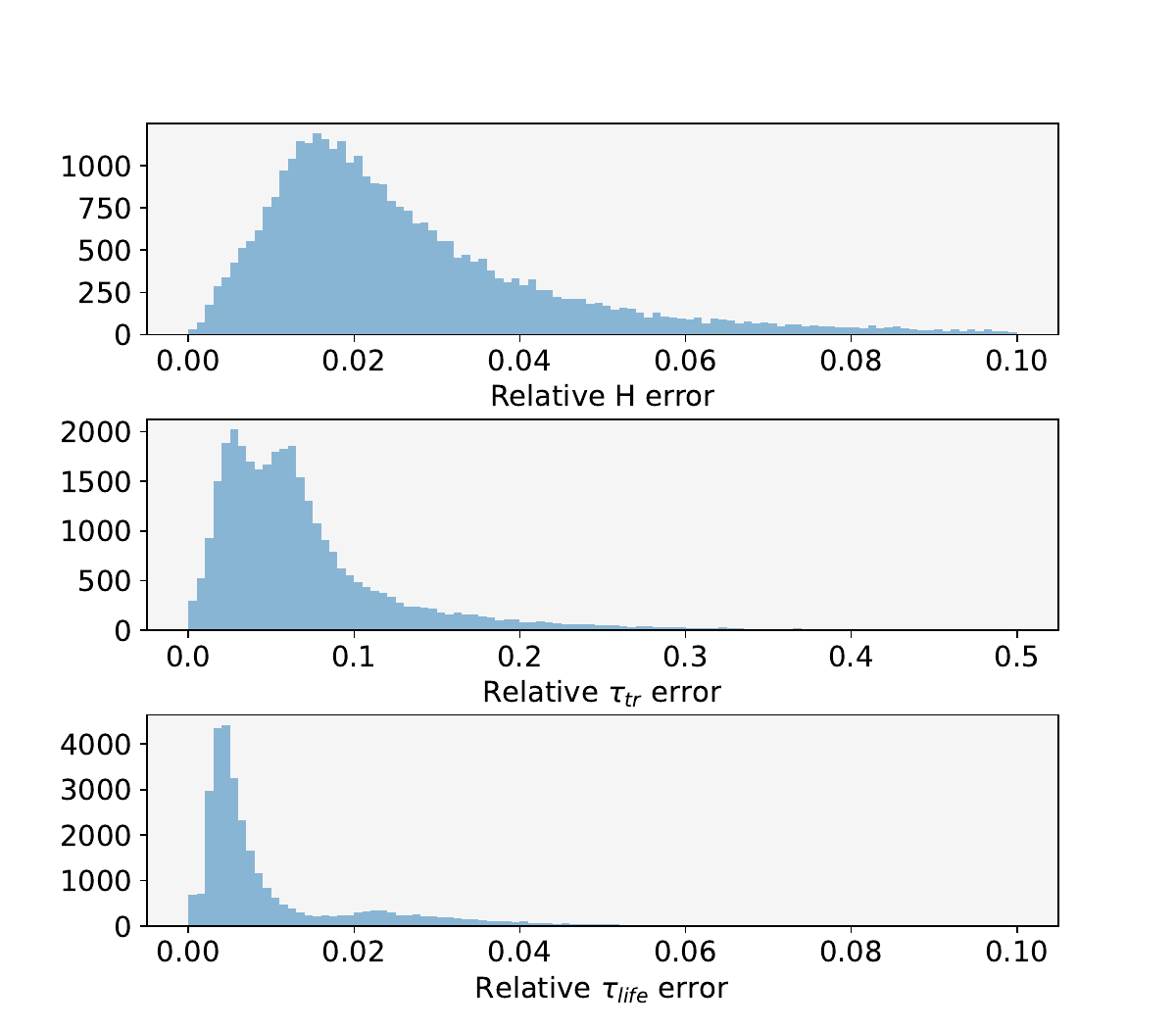}}

      \caption{Histograms of the relative errors of the fitting algorithm for each proxy: $\lnH$ (top), $\ttransit$ (middle) and $\life$ (bottom) for the McQuillan sample. 
              }
         \label{fig:hist_err}
   \end{figure}

Fig. \ref{fig:hist_err} shows the distribution of the relative error estimates obtained using the method of \cite{Toutain1994} for the three parameters to fit in the new model. 
In the case of $\life$, there are two accumulations, one around $0.5\%$ and one around $2.5\%$. As observed with the simulations, this second peak associated with relatively large uncertainty corresponds to the single transit spots case where the rotation peaks are reduced or not apparent, and where the fit has more difficulty converging.  
To avoid spurious estimates, we choose to remove the data with a relative error higher than $10\%$ for $\lnH$ and $\life$, and $20\%$ for $\ttransit$, 
which are
not essential for the present study. 
By adding these criteria, $8.5\%$ and $18.2\%$ of the McQuillan's and Santos' sets respectively are removed. 
We also remove some incorrect fit results by visual inspection of dubious points accumulation in theresults (more than 10\,000 stars). 
The final set after this cleaning consist of approximately 27\,000 stars, which remains a significant sample, and reassures us in the ability of this study to extract global trends.

For the Harvey profile, the same criteria have been applied to $\lnH$ and for the characteristic time; we kept the value with a relative error under 20 $\%$. 
Finally, 5 $\%$ have been removed from the Santos data and 2.6 $\%$ from the McQuillan sample, leading to a total cleaned sample of 41\,176 stars.

The reason we obtain convergence of the parameters on less data with our model is that it depends on 7 free parameters against 4 for the Harvey profile, which makes the algorithm less robust.

After sorting the spectra, we identified 15 092 stars with peakless spectra and 18 663 with rotation peaks, for a total sample of 33\,755 stars. 

\section{Transit proxy as a sanity check}
\label{appendix:time}

 \begin{figure}
    \centering
    \includegraphics[width=\hsize]{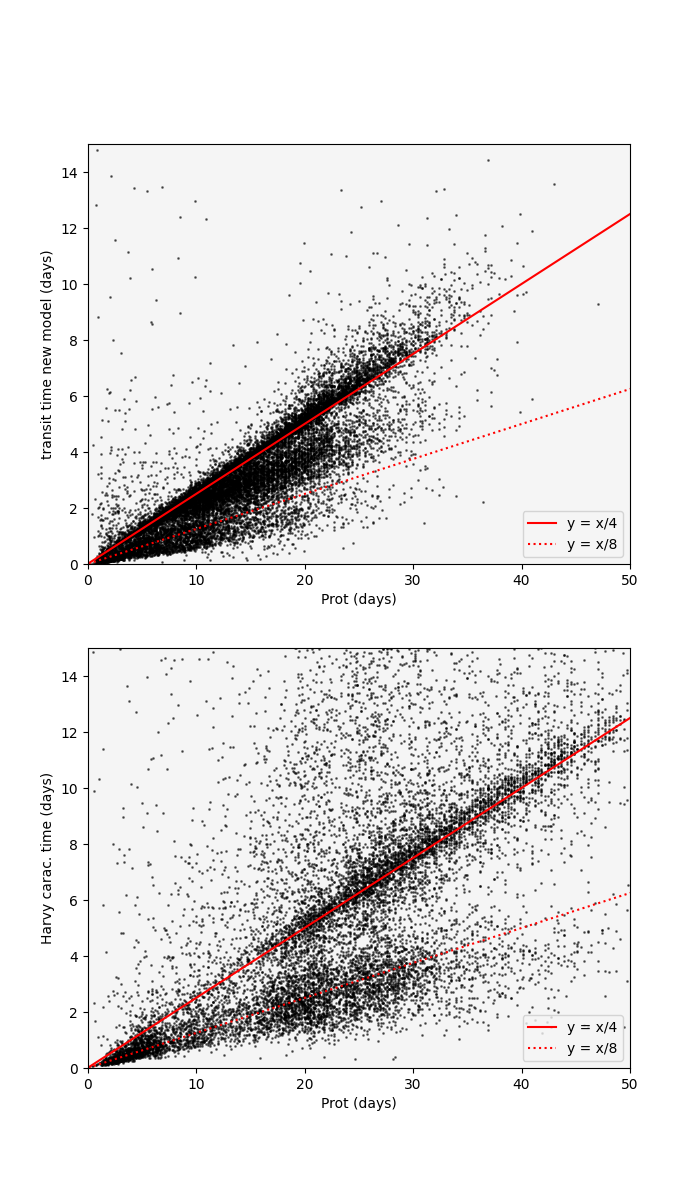}
       \caption{Top panel: transit proxy $\ttransit$ as a function of the rotation period, estimated for the whole stars sample. The red line shows the relation $\ttransit = \prot \//4 $, and the dotted red line its harmonic $\ttransit = \prot \//8$.
       Bottom panel: Harvey characteristic time as a function of the rotation period.  }
          \label{fig:tau_transit}
    \end{figure}

The top panel of Fig.\,\ref{fig:tau_transit} shows the results for the transit proxy as a function of the rotation period estimated by \citet{McQ2014}. 
There is a first linear distribution around the relation, $\ttransit = \prot \//4$, which corresponds to the prediction of the model detailed in Section\,\ref{sec2}.
This distribution is also clearly shifted downwards.
 
The accumulation of points around the red dotted line consist of a set of data for which the algorithm has not converged well.
For example, in some cases there are more than two rotation peaks or other artefacts at higher frequencies, which may bias the estimation of the transit time and lead to an underestimation. In other cases, the first harmonic has very low amplitude, and is not taken into account by the fitting algorithm, which biases the value, as predicted by the simulations.  
Another bias can be an incorrect estimation of the rotation period by \cite{McQ2014} or \cite{Santos2019} by a factor of two, which can occurs with the methods used in these papers. 
Since this $\ttransit$ is not a parameter of interest in this paper, these values have been retained. 

The top panel of Fig.\,\ref{fig:tau_transit} shows the same figure but with the times estimate by the Harvey profile. The results are significantly the same, which reinforces the idea that this parameter is close to the transit time.

\end{appendix}

\end{document}